\documentclass[pra,twocolumn,longbibliography]{revtex4-2}
\usepackage[T1]{fontenc}
\usepackage{amsmath}
\usepackage{amssymb}
\usepackage{mathtools}
\usepackage[normalem]{ulem}
\usepackage{graphicx}
\usepackage{siunitx}
\usepackage{mhchem}
\usepackage{datetime}
\usepackage{color}
\usepackage{newtxtext,newtxmath}
\usepackage{microtype}
\usepackage{braket}
\usepackage{xr}
\usepackage{natbib}
\usepackage{hyperref}
\hypersetup{%
    bookmarksopen=false,
    bookmarksnumbered=true,
    pdfnewwindow=true,
    unicode=false,
    colorlinks=true,%
    citecolor=blue,
    linkcolor=black,
    urlcolor=blue,
    filecolor=blue
    }

\renewcommand{\AA}{\mathscr{A}}
\newcommand{\C}{\mathbb{C}}
\newcommand{\del}{\partial}
\newcommand{\eps}{\varepsilon}
\renewcommand{\i}{\mathrm{i}}
\newcommand{\II}{\mathscr{I}}
\renewcommand{\Im}{\operatorname{Im}}
\newcommand{\KK}{\mathcal{K}}
\newcommand{\LL}{\mathcal{L}}
\renewcommand{\O}{\mathcal{O}}

\newcommand{\RR}{\mathscr{R}}
\renewcommand{\Re}{\operatorname{Re}}
\newcommand{\Tr}{\operatorname{Tr}}
\renewcommand{\vec}[1]{\mathbf{#1}}
\newcommand{\Z}{\mathbb{Z}}
\begin{document}
\newcommand{\figdir}{.}
\newcommand{\liege}{Institut de Physique Nucléaire, Atomique et de Spectroscopie, CESAM, Universit\'e de Li\`ege, 4000 Liège, Belgium}
\title{Dissipative phase transition: from qubits to qudits}

\author{Lukas Pausch}
\altaffiliation[Present address: ]{German Aerospace Center, Institute of Quantum Technologies, Ulm, Germany}
\email{lukas.pausch@dlr.de}
\affiliation{\liege}
\author{François Damanet}
\affiliation{\liege}
\author{Thierry Bastin}
\affiliation{\liege}
\author{John Martin}
\email{jmartin@uliege.be}
\affiliation{\liege}

\begin{abstract}
	We investigate the fate of dissipative phase transitions in quantum many-body systems when the individual constituents are qudits ($d$-level systems) instead of qubits.
	As an example system, we employ a permutation-invariant $XY$ model of $N$ infinite-range interacting $d$-level spins undergoing individual and collective dissipation.
	In the mean-field limit, we identify a dissipative phase transition, whose critical point is independent of $d$ after a suitable rescaling of parameters.
	When the decay rates between all adjacent levels are identical and $d\geq 4$, 
    the critical point expands, in terms of the ratio between dissipation and interaction strengths, to a critical region in which two phases coexist and which increases as $d$ grows.
	In addition, a larger $d$ leads to a more pronounced change in spin expectation values at the critical point.
	Numerical investigations for finite $N$ reveal symmetry breaking signatures in the Liouvillian spectrum at the phase transition.
	The phase transition is furthermore marked by maximum entanglement negativity and a significant purity change of the steady state, which become more pronounced as $d$ increases.
	Considering qudits instead of qubits thus opens new perspectives on accessing rich phase diagrams in open many-body systems.
\end{abstract}
\maketitle

\section{Introduction}

In quantum systems composed of many identical constituents, the interplay of interactions, driving and dissipation gives rise to various collective effects, such as sub- and superradiance \cite{Dicke1954,Crubellier1986,Dimer2007,Lin2012,Hebenstreit2017,Shammah2018, RubiesBigorda2022,Suarez2022}, 
spin squeezing \cite{WinelandSpinSqueezing1992,KitagawaSpinSqueezing1993, MaSpinSqueezing,Norris2012,Lee2014,PezzeSpinSqueezing,Shammah2018,Tucker2020}, or dissipative time crystals \cite{Iemini2018,Shammah2018,Buca2019,
	SachaBook,HuybrechtsThesis,
	Kongkhambut2022,SilvaSouza2023,Mattes2023,Cabot2023,Cabot2024},
with dissipative phase transitions \cite{Lee2014,Minganti2018,Huybrechts2019,Huber2020,Huybrechts2020,Huybrechts2020b,Minganti2021,HuybrechtsThesis,Debecker2023,Wang2023e} connecting phases of distinct steady-state properties. 
With the emergence of experimental platforms such as cold atoms in cavities and optical lattices \cite{RodriguezChiacchio2018,Samutpraphoot2020},
Rydberg atoms \cite{Nill2022,Suarez2022} or trapped ions \cite{Molmer1999,Lin2013,Katz2022} 
that can now be implemented in laboratories with a high degree of control, these fundamental physical phenomena can be studied in detail to determine, among other things, what distinguishes them from their classical counterparts.

Besides bosons \cite{Poletti2012,RodriguezChiacchio2018,Vicentini2018,Giraldo2020,Huybrechts2020b,HuybrechtsThesis,Minganti2022b}, in most theoretical models for 
dissipative interacting many-particle systems 
the individual particles are two-level systems (qubits), i.e., the simplest quantum systems possible. 
For these types of models, efficient numerical methods exist that can simulate the dynamics of 100 qubits and more \cite{Chase2008,Shammah2018}.
However, most physical systems treated as qubits actually host more than two levels and their multilevel nature gives rise to 
effects not explained by a qubit model 
\cite{Lin2012,Kaur2021,HuybrechtsThesis,Suarez2023}.
Indeed, multilevel quantum systems (qudits) offer various advantages over qubits, such as 
larger information capacity \cite{Luo2014a,Cozzolino2019,Wang2020},
more efficient implementations of quantum gates and algorithms \cite{Lanyon2008,Luo2014,Luo2014a,Bocharov2017,Babazadeh2017,Lu2019,Wang2020},
and also of quantum simulation schemes \cite{Neeley2009},
a better protection against noise \cite{Liu2009,Ecker2019,Srivastav2022},
increased security in quantum key distribution \cite{BechmannPasquinucci2000,Cerf2002,Huber2013,Sheridan2010} and quantum communication \cite{Cozzolino2019},
more efficient quantum error correction schemes \cite{Gottesman2001,Cafaro2012,Muralidharan2017,Grassl2018,LoPiparo2020,Lim2023},
and an enhanced sensitivity for 
quantum imaging \cite{Lloyd2008} and 
quantum metrology \cite{Fickler2012,Bouchard2017,Shlyakhov2018}.
Implementations of qudits include a variety of systems, such as photons \cite{Lanyon2008,Lloyd2008,Fickler2012,Bouchard2017,Cozzolino2019,Ecker2019,Gao2020}, 
ultracold atoms \cite{Kasper2021,Dong2023},
trapped ions \cite{Klimov2003,Low2020,Ringbauer2022,Nikolaeva2024},
Rydberg atoms \cite{Ahn2000}, 
nuclear spins \cite{Dogra2014,Gedik2015},
superconducting devices \cite{Neeley2009,Kiktenko2015,Kiktenko2015a,Shlyakhov2018,Kaur2021,CerveraLierta2022,Fischer2023,Subramanian2023} and solid-state defects such as nitrogen-vacancy centers
in diamond
\cite{Jelezko2006,Doherty2013}. 
Furthermore, qudits can emerge as an effective description of bosonic systems, when the system parameters allow a truncation of the number of excitations \cite{Hartmann2010,Finazzi2015}.
Given the advantages of qudits for quantum information tasks and their presence in many different physical platforms, it is rather natural to investigate qudit models also for collective effects in dissipative many-particle systems.

In this work, we explore a multilevel generalisation of the dissipative Lipkin-Meshkov-Glick model~\cite{Lipkin1965,Meshkov1965,Glick1965}. For qubits, this model exhibits a dissipative phase transition between a symmetric phase and a broken-symmetry phase, as the ratio of interaction and (individual or collective) decay is varied~\cite{Lee2013,Joshi2013,Lee2014}. 
We go beyond these known results and examine the presence of a transition, its nature and its characteristics also for $d>2$ levels per constituent, employing both a mean-field analytical approach and finite-size numerics~\cite{Gegg2016,GeggThesis,HuybrechtsThesis,Sukharnikov2023}.
For all $d$, a dissipative phase transition arises. 
Through a suitable rescaling of the decay rates, the position of the critical point and the qualitative characteristics of the two phases can be made insensitive to the number of single-particle levels and to
the exact nature of the decay, which can here be modelled more flexibly than for two-level systems. 
For specific choices of the decay, which are accessible only for $d\geq 4$, the critical point evolves into a critical region, whose size increases with $d$. 
Furthermore, we show that the Liouvillian spectral properties, and steady-state spin expectation values, purity and entanglement become more sensitive indicators of the phase transition as the number of levels per particle increases.

This manuscript is organized as follows: After presenting the model for a general number of single-particle levels in Section~\ref{sec:Model}, we consider, in Section~\ref{sec:mean-field}, its mean-field limit to discuss the properties of its dissipative phase transition
 for qubits and for qudits, 
 highlighting the differences between two-level and multilevel systems. Sections~\ref{sec:numericsL} and~\ref{sec:numericsRho} are devoted to numerical results for the Liouvillian spectrum and for the steady-state properties at finite numbers of particles. 
In Section~\ref{sec:Experiment}, we sketch a possible experimental implementation of the model, 
before we conclude in Section~\ref{sec:conclusions}.

\section{Model and its symmetries}
\label{sec:Model}

\begin{figure}
	\includegraphics{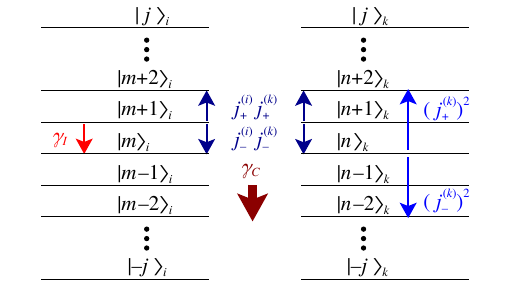}
	\caption{Effects of the model Hamiltonian [Eq.~\eqref{eq:Hamiltonian}, blue] and the dissipators [Eq.~\eqref{eq:dissipator}, red] on two spin-$j$ particles labelled by $i$ and $k$, with single-particle levels $\ket{m}_i$, $\ket{n}_k$ ($m,n = -j,\ldots,j$) such that $j_z^{(i)}\ket{m}_i = m \ket{m}_i$ and $j_z^{(k)} \ket{n}_k = n \ket{n}_k$.
	We choose different symbols $m$, $n$ for the quantum numbers of particles $i$, $k$ to stress that the two-body interaction terms $j_\pm^{(i)} j_\pm^{(k)}$ act on all states $\ket{m}_i\otimes\ket{n}_k$, where $m$ and $n$ need not be identical.
	}
	\label{fig:SketchModel}
\end{figure}

We consider $N$ identical spins, each with $d=2j+1$ levels, where $j$ is the spin quantum number.
In the single-particle basis $\ket{m}_i$ ($i=1,\ldots,N$, $m=-j,\ldots,j$), the individual spin operators of the $i$th particle are~\cite{Sakurai}
\begin{subequations}
\begin{align}
	& j_+^{(i)} = \sum_{m=-j}^{j-1} A_{j,m}\ket{m+1}_i\bra{m}_i, \quad j_-^{(i)} = (j_+^{(i)})^\dagger,\\
	& j_x^{(i)} = \frac{1}{2}\left(j_+^{(i)} + j_-^{(i)}\right),\quad j_y^{(i)} = \frac{1}{2\i}\left(j_+^{(i)} - j_-^{(i)}\right),\\
	& j_z^{(i)} = \sum_{m=-j}^{j} m \ket{m}_i\bra{m}_i,
\end{align}
\end{subequations}
with $A_{j,m} = \sqrt{(j-m)(j+m+1)}$, 
and the collective spin operators are constructed as $J_\alpha = \sum_{i=1}^{N} j_\alpha^{(i)}$, $\alpha = \pm,x,y,z$. 
The unitary dynamics of the spins is assumed to be given by the Hamiltonian~\cite{Lee2014}
\begin{align}
	H = \frac{V}{Nj} \left( J_x^2 - J_y^2 \right) = \frac{V}{2 N j}\left( J_+^2 + J_-^2 \right).
	\label{eq:Hamiltonian}
\end{align}
This model is a special case of the Lipkin-Meshkov-Glick (LMG) model~\cite{Lipkin1965,Meshkov1965,Glick1965}, with maximal anisotropy between $J_x^2$ and $J_y^2$ and no effective magnetic field in the $z$ direction.
It can also be understood as an $XY$ model \cite{Lieb1961,Barouch1970} with infinite-range interactions. 
For any $d$, the Hamiltonian contains a two-body interaction term $\sim \sum_{i=1}^{N} \sum_{k=i+1}^{N} \left(j_+^{(i)} j_+^{(k)} + j_-^{(i)} j_-^{(k)}\right)$ leading to the joint excitation or deexcitation of two spins. For $d>2$, i.e., $j> 1/2$,
 the Hamiltonian furthermore contains a single-particle term $\sim \sum_{i=1}^{N} \left( (j_+^{(i)})^2 + (j_-^{(i)})^2 \right)$, which (de)excites each spin by two levels. 
Figure~\ref{fig:SketchModel} shows a simplified sketch of these terms and of their effect onto the single-particle levels (blue arrows and labels).
Note that the $d$-level nature of the spins would allow us to define also a more general LMG Hamiltonian than the one studied here, whose spectrum and eigenstates have been investigated, e.g., in Refs.~\cite{Gilmore1979,Meredith1988,Wang1998,Gnutzmann1998,Gnutzmann1999,Calixto2021,Calixto2021a}. 

We study the steady-state properties of this model subject to individual and collective spontaneous decay of the spins, described by the following Lindblad master equation for the density matrix $\rho$ (with $\hbar=1$):
\begin{align}
	\nonumber
	\dot{\rho} ={}& -\i [H,\rho] + \frac{\gamma_I}{j}\sum_{i=1}^N \left(L^{(i)}\rho \left(L^{(i)}\right)^\dagger - \frac{1}{2}\left\{\left(L^{(i)}\right)^\dagger L^{(i)}, \rho\right\}\right) \\
	&+ \frac{\gamma_C}{Nj} \left(L_C\rho L_C^\dagger - \frac{1}{2}\left\{L_C^\dagger L_C, \rho\right\}\right),
	\label{eq:master-eq}
\end{align}
with decay rates $\gamma_I$ for individual dissipation and $\gamma_C$ for collective dissipation, and corresponding Lindblad operators $L^{(i)}$, $L_C$.
Specifically, we consider
\begin{align}
	L^{(i)} = \sum_{m=-j}^{j-1} \ell_m \ket{m}_i\bra{m+1}_i
	\label{eq:dissipator}
\end{align}
and $L_C = \sum_{i=1}^{N} L^{(i)}$, i.e., $d$-level generalizations of the two-level dissipators $L^{(i)} = \sigma_-^{(i)}$ and $L_C = J_-$. 
This kind of collective dissipation, where $L_C$ is the sum of identical individual dissipators, naturally arises in effective models of $N$ atoms coupled identically to a collection of modes in a dissipative cavity, when the cavity dynamics can adiabatically be eliminated \cite{Morrison2008,Morrison2008a,Huber2020}, as we discuss also in Sec.~\ref{sec:Experiment}.
Clearly, the operators $L^{(i)}$ that define $L_C$ might differ from those that describe the individual dissipation. However, as we will outline in Sec.~\ref{sec:Experiment}, the collective dissipator $L_C$ can be implemented experimentally in a tunable way and thus be chosen as the sum of exactly the same $L^{(i)}$ that also define the individual dissipation.

A simplified sketch of the effect of such operators $L^{(i)}$ and $L_C$
is shown in Fig.~\ref{fig:SketchModel} (red arrows and labels): 
the individual dissipator $L^{(i)}$ corresponds to quantum jumps of particle $i$ from $\ket{m+1}_i$ to $\ket{m}_i$ and the collective dissipator $L_C$ gives rise to quantum jumps from collective states with $N_{m+1}$ and $N_m$ particles in the single-particle states~$\ket{m+1}$,~$\ket{m}$, respectively, to collective states with $N_{m+1}-1$ and $N_m+1$ particles in these two states.
Such dissipation processes typically arise when the system is coupled to a zero-temperature bath that drives the system into its ground state. With the Hamiltonian~\eqref{eq:Hamiltonian} considered here, however, the processes described by the jump operators do not necessarily lower the energy. This apparent contradiction is resolved in the possible experimental implementation that we propose in Section~\ref{sec:Experiment}: There, the Hamiltonian~\eqref{eq:Hamiltonian} arises as an effective Hamiltonian in a driven, i.e., time-dependent system, whereas the dissipation processes indeed lead to decays from higher to lower energy in the original, undriven system.

Jointly scaling $\ell_m \mapsto c \ell_m$ ($m=-j, \ldots,j-1$) and $\gamma_{k} \mapsto |c|^{-2} \gamma_{k}$ ($k=I,C$) with $c \in \C $ leaves the master equation invariant. As a convention, we fix the decay rates and Lindblad operators such that $\ell_{-j} = \sqrt{2 j}$.
Examples of such operators are the spin ladder operator 
\begin{align}
	L^{(i)}_{\text{spin}} = j_{-}^{(i)},
	\label{eq:spin-ladder}
\end{align}
with $\ell_m = A_{j,m}$ ($m=-j,\ldots,j-1$) and $L_C = J_-$, and the $m$-independent dissipator 
\begin{align}
	L^{(i)}_{\equiv} = \sqrt{2j}\sum_{m=-j}^{j-1} \ket{m}_i\bra{m+1}_i.
	\label{eq:m-independent}
\end{align}
Note that for $d=2$ and $d=3$, the latter two individual dissipators \eqref{eq:spin-ladder} and \eqref{eq:m-independent} are identical to each other, but they differ for $d\geq 4$. 

For qubits ($d=2$), this model~\cite{Lee2013,Lee2014} and 
variants of it~\cite{Joshi2013}
have been studied previously with respect to their steady-state properties and dissipative phase transitions in the limit $N\to\infty$. Furthermore, variants including more general collective dissipation, but without individual dissipation, have been considered in the limit $Nj\to\infty$~\cite{Ferreira2019}. 

The right-hand side of
Eq.~\eqref{eq:master-eq} is 
invariant under permutations of the 
particles. 
Hence, a permutation-invariant density matrix $\rho$, i.e., 
$[\rho,\pi]=0$ for any $N$-particle permutation operator $\pi$, 
stays permutation-invariant throughout the whole time evolution provided by Eq.~\eqref{eq:master-eq} \cite{Chase2008,Shammah2018,HuybrechtsThesis}. 

In addition, the master equation obeys a $\Z_2$ symmetry mediated by the unitary superoperator $\Pi_1: A \mapsto e^{\i \pi J_z} A e^{-\i\pi J_z}$ and, if $L^{(i)}$ and $L_C$ are real operators
(as it is the case for $L^{(i)}_{\text{spin}}$ and $L^{(i)}_{\equiv}$), another $\Z_2$ symmetry given by the antiunitary superoperator $\Pi_2: A \mapsto e^{\i \pi J_z/2} A^* e^{-\i\pi J_z/2}$, where $A^*$ is the complex conjugate of $A$.
These symmetries entail that for any steady-state solution $\rho$ of Eq.~\eqref{eq:master-eq}, $\Pi_k[\rho]$ ($k=1,2$) are also steady states. The expectation values of an observable $O$ in these states are related via $\left<O\right>_{\Pi_k[\rho]} = \langle\Pi_k^\dagger[O]\rangle_{\rho}$ (where $\left<O\right>_A = \Tr[OA]$), in particular 
$\left<J_x\right>_{\Pi_1[\rho]} = -\left<J_x\right>_{\rho}$, $\left<J_y\right>_{\Pi_1[\rho]} = -\left<J_y\right>_{\rho}$, $\left<J_z\right>_{\Pi_1[\rho]} = \left<J_z\right>_{\rho}$ and $\left<J_x\right>_{\Pi_2[\rho]} = \left<J_y\right>_{\rho}$, $\left<J_y\right>_{\Pi_2[\rho]} = \left<J_x\right>_{\rho}$, $\left<J_z\right>_{\Pi_2[\rho]} = \left<J_z\right>_{\rho}$.\\

\section{Dynamics and Dissipative Phase Transition in the Thermodynamic Limit}
\label{sec:mean-field}

To get an understanding of the system dynamics, let us first investigate the model in the limit $N\to\infty$. For this purpose, we derive the equations of motion of the following collective Hermitian operators
\begin{subequations}
\label{eq:S-operators}
\begin{align}
	S_{x;m,n} = \frac{1}{2N}\sum_{i=1}^N \left(\ket{m}_i\bra{n}_i + \ket{n}_i\bra{m}_i\right),\\
	S_{y;m,n} = \frac{1}{2\i N}\sum_{i=1}^N \left(\ket{m}_i\bra{n}_i - \ket{n}_i\bra{m}_i\right),
\end{align}
\end{subequations}
where $m,n=-j,\ldots,j$ and by convention $m\geq n$ (note, however, that $m=n$ for $y$ corresponds to $S_{y;m,m}=0$).
Collective spin operators are given in terms of these operators as $J_x/N = \sum_{m=-j}^{j-1} A_{j,m} S_{x;m+1,m}$, $J_y/N = \sum_{m=-j}^{j-1} A_{j,m} S_{y;m+1,m}$, $J_z/N = \sum_{m=-j}^{j} m S_{x;m,m}$.
The expectation values of the $S$ operators are the matrix elements of the averaged single-particle density matrix $\bar{\rho} = \sum_{i=1}^N \rho^{(i)}/N$,
where $\rho^{(i)}$ is obtained from the full density matrix $\rho$ by taking the trace over all particles but the $i$th one ($i=1,\ldots,N$):
\begin{align}
	\left<S_{x;m,n}\right> &= \Re [\bar{\rho}_{nm}], &
	\left<S_{y;m,n}\right> &= \Im [\bar{\rho}_{nm}].
\end{align}
Consequently, $\sum_{m=-j}^{j} \left<S_{x;m,m}\right> = \Tr[\bar{\rho}] = 1$ and $\sum_{m,n = -j}^j\left<S_{x;m,n}\right>^2 + \left<S_{y;m,n}\right>^2 = \Tr[\bar{\rho}^2] \leq 1$.

Assuming that the expectation values of the individual terms $\ket{m}_i\bra{n}_i$ etc. are independent of $N$ and $i$ for $N\to\infty$, 
one can easily check that $\left<S_{\alpha;m,n}\right>$ ($\alpha =x,y$) scales as $\O(1)$, whereas the expectation value of the commutator $[S_{\alpha;m,n},S_{\beta;o,p}]$ ($\alpha,\beta =x,y$) scales as $\O(N^{-1})$ and hence vanishes as $N\to\infty$. 
This justifies treating the $S$ operators as real numbers in that limit and in particular assuming $\left<\left\{S_{\alpha;m,n}, S_{\beta;o,p}\right\}\right> - 2\left<S_{\alpha;m,n}\right> \left<S_{\beta;o,p}\right> \to 0$ for $N\to\infty$,
an assumption whose validity can also be proved analytically for rather general cases including the $\gamma_C=0$ and the $\gamma_I=0$ limits of our model \cite{Fiorelli2023,Carollo2024}.
Equations of motion for the expectation values in the limit $N\to\infty$ then become (for further details see Appendix~\ref{sec:details-mean-field}) 
\begin{align}
	\nonumber
	 \left<\dot{S}_{\alpha;m,n}\right> ={}& V 
	\;f_{\alpha;m,n}\!\left(\left<S_{\beta;o,p}\right>\right) + \gamma_I \; g_{\alpha;m,n}\!\left( \left<S_{\beta;o,p}\right>\right) \\
	&+ \gamma_C \; h_{\alpha;m,n}\!\left(\left<S_{\beta;o,p}\right>\right),
	\label{eq:mean-field}
\end{align}
where $\alpha,\beta=x,y$ and $m,n,o,p =-j,\ldots,j$,
and where $f_{\alpha;m,n}$, $h_{\alpha;m,n}$ are two different quadratic functions and $g_{\alpha;m,n}$ a linear function of the $\left<S_{\beta;o,p}\right>$. The coefficients of $f_{\alpha;m,n}$ are of the form $A_{j,\mu} A_{j,\nu}$, whereas the coefficients of $g_{\alpha;m,n}$ and $h_{\alpha;m,n}$ are of the forms $\Re[\ell_\mu^*\ell_\nu]$, $\Im[\ell_\mu^*\ell_\nu]$ and $|\ell_{\mu}|^2+|\ell_{\nu}|^2$ ($\mu,\nu = -j,\ldots,j-1$).
The steady states of this set of equations are the stable fixed points, i.e., solutions to 
$\left<\dot{S}_{\alpha;m,n}\right>(t) = 0$
such that 
the Jacobian matrix, i.e., the gradient of the right-hand side of Eq.~\eqref{eq:mean-field}, has only eigenvalues with negative real part \cite{Ott2002,Strogatz2015}.

	\subsection{Qubits}
	\label{sec:Qubits}
	
	Let us first 
	recapitulate
	previously obtained results for the qubit LMG model \cite{Lee2013,Lee2014}.
	In this case, 
	the system dynamics can conveniently be described in terms of scaled spin expectation values $X = \left<J_x\right>/N j = 2\left<S_{x;1/2,-1/2}\right>$, $Y= \left<J_y\right>/N j= 2\left<S_{y;1/2,-1/2}\right>$, $Z= \left<J_z\right>/N j = \left<S_{x;1/2,1/2}\right> - \left<S_{x;-1/2,-1/2}\right>$. 
	Equation~\eqref{eq:mean-field} then reduces to~\cite{Lee2013,Lee2014}
	\begin{subequations}
	\label{eq:qubits}
	\begin{align}
		\dot{X} &= - 2 V Y Z - \gamma_I X + \gamma_C X Z, \\
		\dot{Y} &= - 2 V X Z - \gamma_I Y + \gamma_C Y Z, \\
		\dot{Z} &= 4 V X Y - 2 \gamma_I (Z+1) - \gamma_C (X^2 + Y^2).
	\end{align}
	\end{subequations}
	For all $V$, $\gamma_I$, $\gamma_C$, the point $X=Y=0$, $Z=-1$ is a fixed point, 
	which is stable for $\gamma_I + \gamma_C > 2|V|$.
	This state, which we may call the spin-$z$ polarized steady state, is invariant under both symmetries $\Pi_1$ and $\Pi_2$.
	For $\gamma_I + \gamma_C < 2|V|$ 
	and $\gamma_I\neq 0$, two different 
	steady states 
	emerge at $X = \operatorname{sgn}(V) Y = \pm \sqrt{\gamma_I\left(2|V| - \gamma_I - \gamma_C\right)}/\left(2|V|-\gamma_C\right)$, $Z = - \gamma_I/\left(2|V|-\gamma_C\right)$. Since these two states do not obey $\Pi_1$ symmetry (instead, they are mapped to each other by $\Pi_1$), we may call them the 
	broken-symmetry
	steady states. The steady-state coordinates $X$, $Y$, $Z$ are continuous at $\gamma_I + \gamma_C = 2|V|$, while their first derivative with respect to $\gamma_{k}/V$ ($k=I,C$) is not, i.e., this dissipative phase transition is a second-order transition \cite{Lee2013,Lee2014}.
	
	If $\gamma_I = 0$, the quantity $r=\sqrt{X^2+Y^2+Z^2}$, i.e., the length of the total spin vector,
	is conserved and can be fixed at its maximal value $r=1$.
	The unique steady state for 
	$\gamma_C > 2|V|$
	is $X=Y=0$, $Z=-1$ like for $\gamma_I\neq 0$~\cite{Lee2014}\footnote{Additionally, for $\gamma_I=0$ an unstable fixed point is found at $X=Y=0$, $Z=+1$}. For 
	$\gamma_C < 2|V|$, 
	this state becomes unstable, and four other fixed points emerge~\cite{Lee2014}, the two points
	\begin{subequations}
	\label{eq:steady-state-qubits-collective}
	\begin{align}
		X &= \sqrt{\frac{1}{2} \pm \frac{1}{2}\sqrt{1-\left(\frac{\gamma_C}{2V}\right)^2}},\\
		Y &= \operatorname{sgn}(V)\sqrt{\frac{1}{2} \mp \frac{1}{2}\sqrt{1-\left(\frac{ \gamma_C}{2V}\right)^2}},\\
		Z &= 0,
	\end{align}
	\end{subequations}
	and the two points obtained from them via the mapping $(X,Y)\mapsto(-X,-Y)$, i.e., the symmetry operation $\Pi_1$. 
	The eigenvalues of the Jacobian at these four points are purely imaginary,
	which means that the steady-state solutions of Eq.~\eqref{eq:qubits} are periodic orbits around the fixed points. The time averages along these orbits fulfill $\overline{X(t)}\neq 0$, $\overline{Y(t)}\neq 0$, $\overline{Z(t)} = 0$ \cite{Lee2014}. Consequently, this time average is discontinuous at $\gamma_C = 2|V|$ and the phase transition is of first order \cite{Lee2014}, in contrast to the second-order transition for $\gamma_I\neq 0$  (even for infinitesimally small $\gamma_I\gtrsim 0$).

	Remarkably, the steady states at $\gamma_I =0$ and $\gamma_C < 2|V|$ differ strongly from the corresponding steady states at infinitesimally small $\gamma_I\gtrsim 0$: 
	In the first case the spin length $r=\sqrt{X^2+Y^2+Z^2}$ is fixed at 1, whereas $r \approx 0$ for the steady states of the second case.
	This strong effect of infinitesimal $\gamma_I$ can be explained from the interplay of the different terms: 
	The time derivative of $r^2$ reveals that individual dissipation reduces the length $r$ as long as 
	$Z+1>\sqrt{1-r^2}$, while the other two terms do not change $r$.
	The oscillations induced by the interaction periodically drive $Z$ above that threshold, such that a small dissipation rate $\gamma_I$ eventually leads to $r \approx 0$. However, the time scales to reach this steady state are expected to diverge as $1/\gamma_I$ for $\gamma_I\to 0$.
	A similarly strong impact of infinitesimally weak individual dissipation has been reported also for the steady state of the Dicke model \cite{Leppenen2024}, where a second-order dissipative phase transition transforms into a first-order transition with a bistable region due to the presence of individual dissipation.
	
	If $\gamma_I$ is larger, the effect of dissipation, which drives the system towards $Z=-1$ (individual dissipation) or $Z=-r$ (collective dissipation), pushes the $Z$ coordinate away from 0 and the steady state is found at a finite $r$, with $Z+1=\sqrt{1-r^2}$. Finally, when the dissipation dominates, the interaction is too weak to lift the system into the parameter region where $r$ is reduced and the steady state obeys $r=1$, $Z=-1$.
	
	\begin{figure}
		\includegraphics{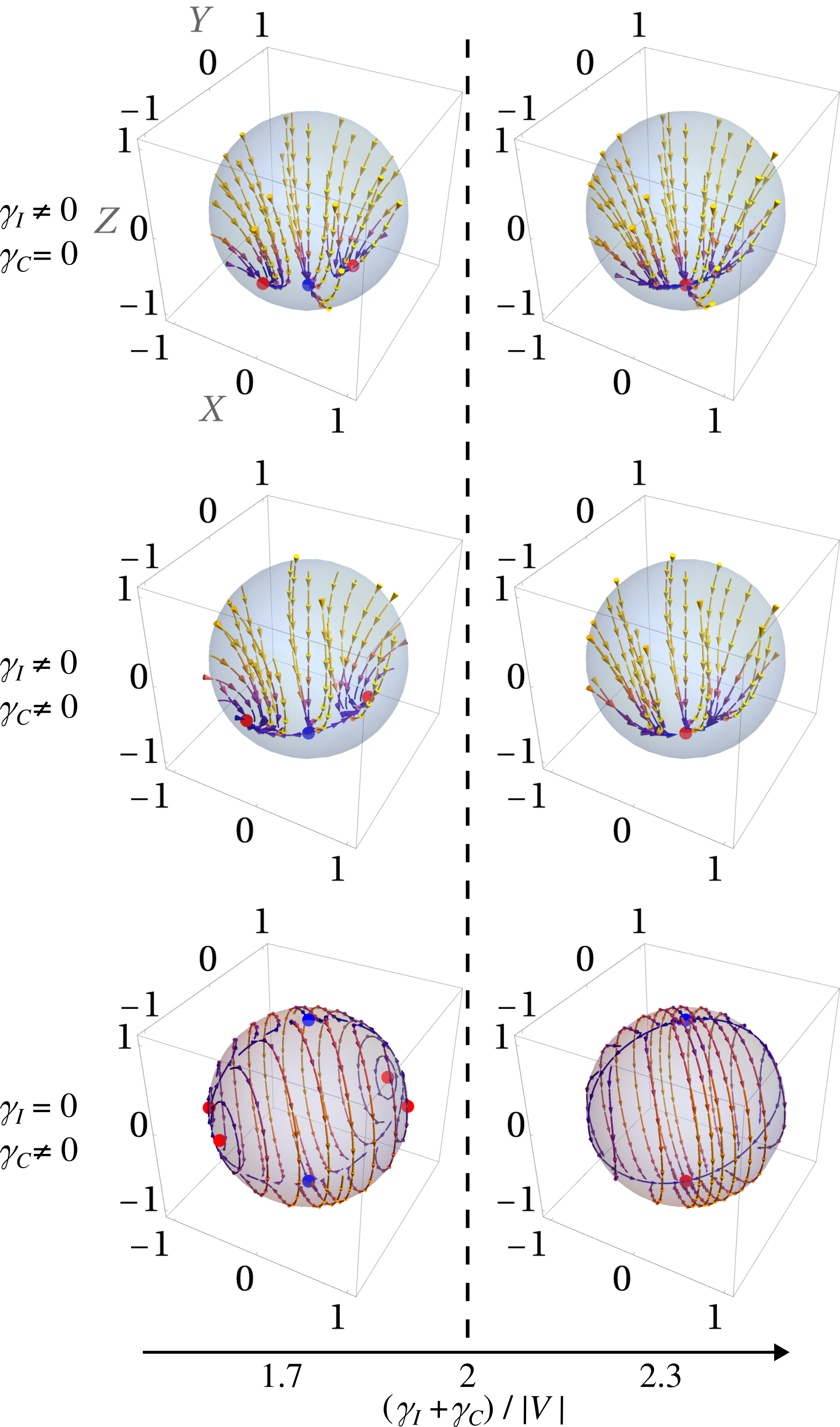}
		\caption{Stream plots of the mean-field equations of motion for qubits, Eq.~\eqref{eq:qubits}, with only individual dissipation (top row), both types of dissipation (middle row, with $\gamma_C = 2\gamma_I$) or only collective dissipation (bottom row), and with ratios $\left(\gamma_{I}+\gamma_C\right)/|V| = 1.7, 2.3$ around the critical value $\left(\gamma_{I}+\gamma_C\right)/|V| = 2$ (dashed line). Here, only $V\geq 0$ is considered. The red (brighter) dots represent the steady states (in the bottom left plot, the center fixed points instead), whereas blue (darker) dots are unstable fixed points. Axes labels of the upper left plot apply to all panels.}
		\label{fig:qubitsMeanfield}
	\end{figure}
	
	The above discussion is illustrated by the streamline plots of Eq.~\eqref{eq:qubits} shown in Fig.~\ref{fig:qubitsMeanfield} for 
	two different ratios $(\gamma_{I}+\gamma_C)/|V|$ and three different values of $(\gamma_I,\gamma_C)$.
	In the dissipation-dominated regime \mbox{$(\gamma_{I}+\gamma_C)/|V|>2$} (right column) all trajectories converge to the unique steady state $(X,Y,Z)=(0,0,-1)$. In contrast, in the interaction-dominated regime \mbox{$(\gamma_{I}+\gamma_C)/|V|<2$} (left column), 
	the behavior depends on whether individual dissipation contributes or only collective dissipation is present: In the former case (top and middle row), two 
	broken-symmetry
	steady states are found, with identical $Z$ value but opposite $X$ and $Y$ values.  
	When both types of dissipation are present (middle row), the 
	broken-symmetry
	fixed points deviate further from the spin-$z$ polarized state
	than for only individual dissipation (top row). 
	In the case of only collective dissipation (bottom row), trajectories describe closed curves on the surface of the sphere and the fixed points are never reached, except for trajectories starting at the \mbox{fixed points themselves}.
	
	\subsection{Qudits}
	
	\label{sec:Qudits}
	
	After having recovered and summarised the results for the qubit LMG model in the preceding section, let us now extend them to general $d > 2$.
	In the following, we will first briefly discuss some analytically accessible results about the steady states for general $d$.
	As a next step, we will numerically investigate in more detail the mean-field dynamics of Eq.~\eqref{eq:mean-field} for $d=4$ and specific values of $\gamma_I$, $\gamma_C$ and $V$, examining the cases of only individual dissipation, both types of dissipation, and only collective dissipation, and highlighting qualitative similarities with $d=2$.
	Finally, we will study the mean-field steady states as a function of $\gamma_I$ for several $d$, which reveals quantitative deviations from $d=2$.
	
	Independently of $L^{(i)}$ 
	and for all values of $d$, $V$, $\gamma_{I}$, $\gamma_C$, the point 
	\begin{align}
		\left< S_{x;-j,-j}\right> = 1, \quad \left< S_{\alpha;m,n}\right> = 0, \; (\alpha;m,n) \neq (x;-j,-j)
		\label{eq:southpole}
	\end{align}
	is a fixed point in the limit $N\to\infty$, which is stable if and only if
	$|\ell_{-j}|^2 \left(\gamma_I + \gamma_C\right) > 4j|V|$, as we show in more detail in Appendix~\ref{sec:details-fixed-points}.
	With our scaling convention $\ell_{-j} = \sqrt{2j}$, 
	the transition point is thus independent of $d$ and coincides with the qubit case, $ \gamma_I + \gamma_C = 2|V|$. The scaled spin expectation values $X = \left<J_x\right>/Nj$, $Y = \left<J_y\right>/Nj$, $Z = \left<J_z\right>/Nj$ (with $N\to\infty$) are also independent of $d$ for this steady state and thus the same as for qubits, $X = Y = 0$, $Z = -1$. We may again refer to that state as the spin-$z$ polarized state.
	
	If there is only collective dissipation (i.e., $\gamma_I=0$), 
	all points with $\left< S_{x;m+1,m}\right> = \left< S_{y;m+1,m}\right> = 0$ for $m=-j,\ldots,j-1$ are moreover fixed points. 
	In addition to 
	$|\ell_{-j}|^2\gamma_C$, $|V|$, and $j$, the stability of these points also depends on the other matrix elements of $L^{(i)}$ and on the differences $\left<S_{x;m+1,m+1}\right>-\left<S_{x;m,m}\right>$ (see Appendix~\ref{sec:details-fixed-points} for more details). 
	Note that conservation of the total spin length $|\vec{J}|$ rules out these fixed points for $d=2$ 
	(except for the 
	spin-$z$ polarized steady state
	and the unstable fixed point with $\left< S_{x;1/2,1/2}\right> = 1$).
	For $d>2$, however, the length of the total spin is not conserved in general and these additional fixed points may be accessed by the system.
	
	Other steady states emerge in the region where the fixed points discussed above become unstable, as was already the case for $d=2$. 
	As these steady states turn out to be analytically accessible only for $d=2$ and $d=3$, 
	we study them numerically and restrict our investigations to the dissipators $L^{(i)}_{\text{spin}} = j_-^{(i)}$ and $L^{(i)}_{\equiv}$ defined in Eqs.~\eqref{eq:spin-ladder} and~\eqref{eq:m-independent}, and corresponding collective dissipators.
	
	\begin{figure*}
		\includegraphics[width=0.49\linewidth]{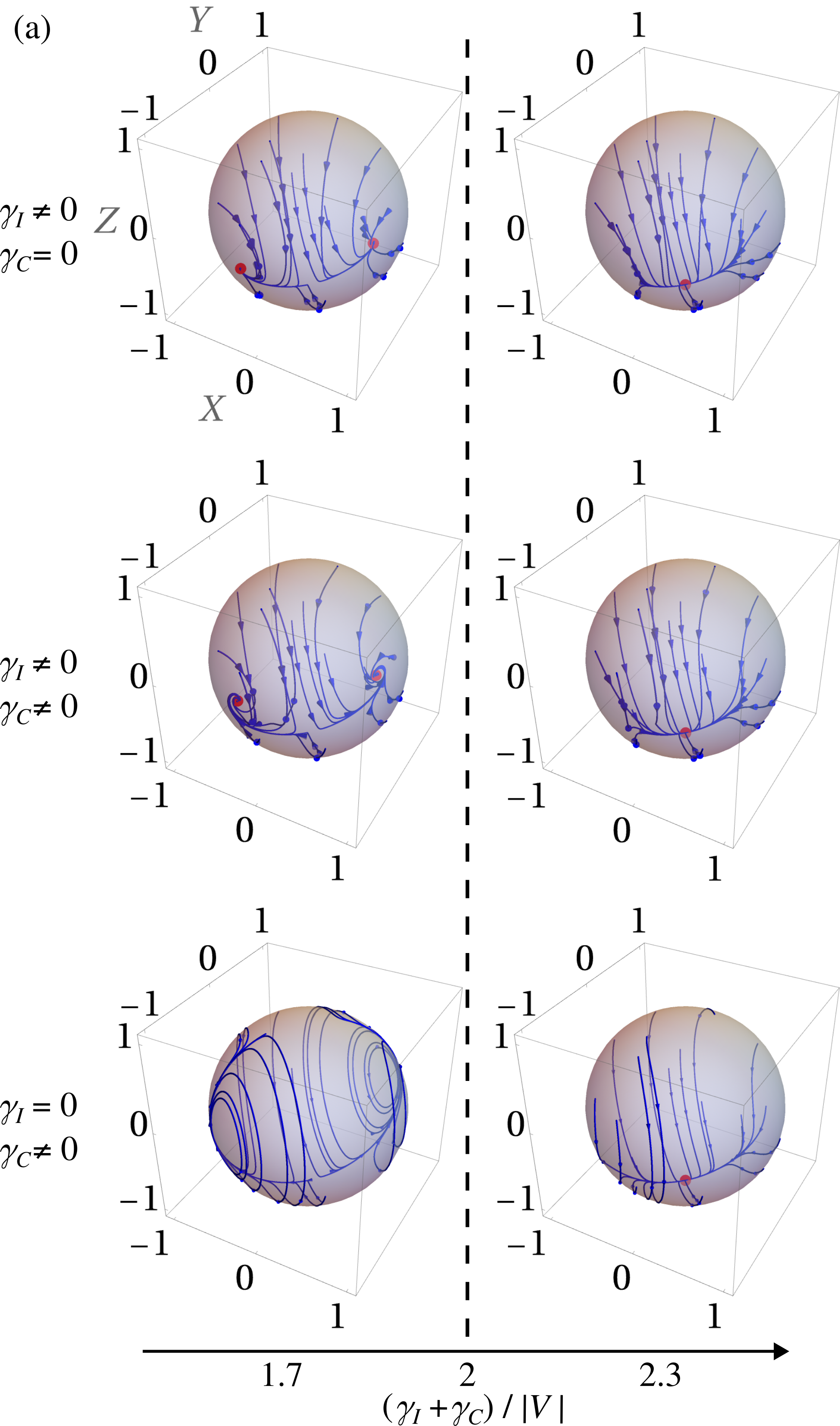}
		\includegraphics[width=0.49\linewidth]{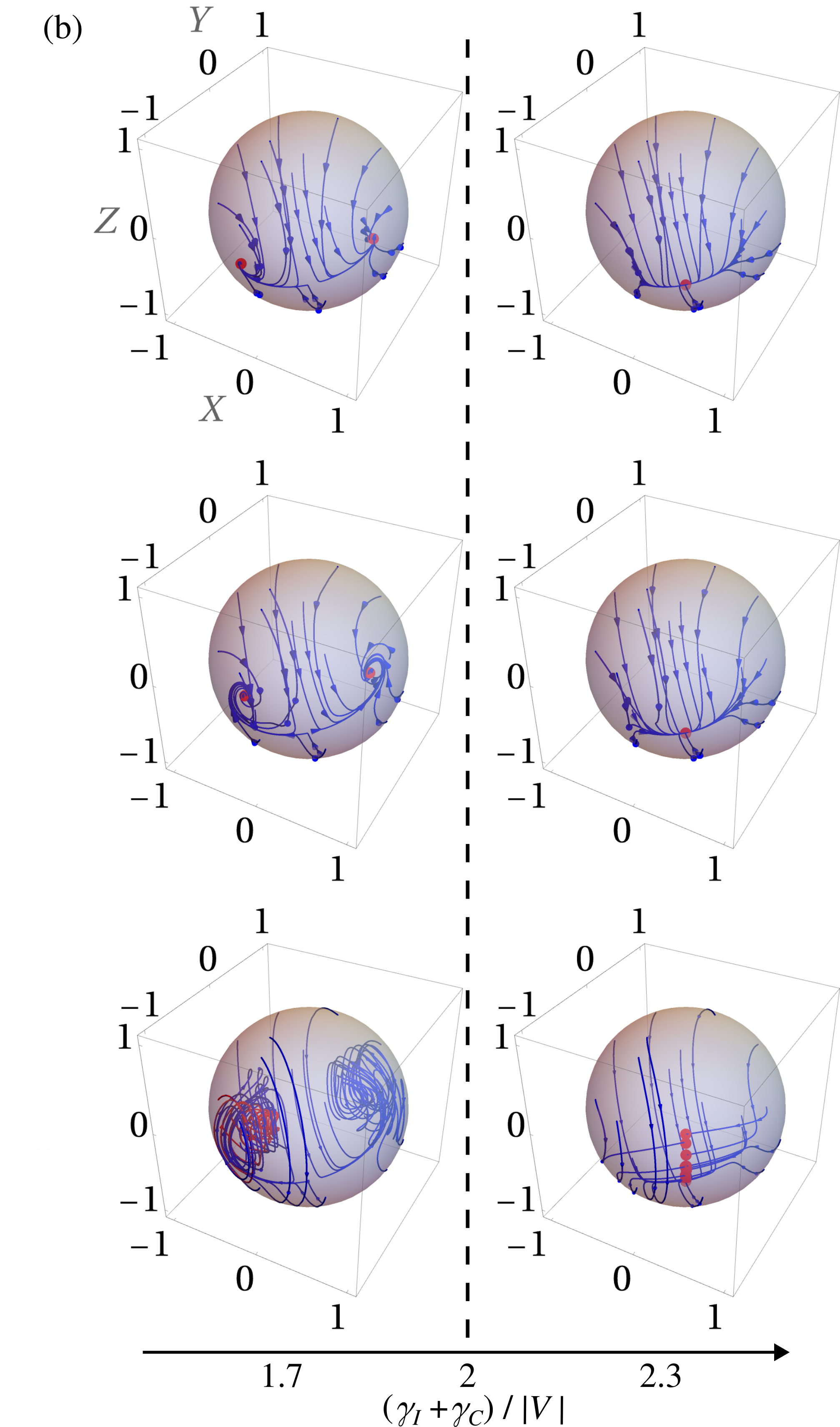}
		\caption{Time evolution of the scaled spin expectation values $X = \left<J_x\right>/Nj$, $Y =  \left<J_y\right>/Nj$, $Z = \left<J_z\right>/Nj$ in the mean-field limit $N\to\infty$ [Eq.~\eqref{eq:mean-field}] for $d=4$, with dissipator $L^{(i)}_{\text{spin}}$ (a) and dissipator $L^{(i)}_{\equiv}$ (b), and only individual dissipation (top row), both types of dissipation (middle row, with $\gamma_C=2\gamma_I$) or only collective dissipation (bottom row), and with ratios $\left(\gamma_{I}+\gamma_C\right)/|V| = 1.7, 2.3$ around the critical value $\left(\gamma_{I}+\gamma_C\right)/|V| = 2$ (dashed line).  Here, only $V\geq 0$ is considered. The 20 initial states of the shown trajectories were chosen randomly from the states with maximal spin length and are the same in all plots. Red dots denote the steady states and the red line (in grayscale, the brighter line at negative $X$ and $Y$) in the lower left plot of (b) highlights a single trajectory referenced in the main text. Axes labels of the upper left plot in each subfigure apply to all panels.}
		\label{fig:dLevelMeanfield_overview}
	\end{figure*}
		
	Figure~\ref{fig:dLevelMeanfield_overview} shows, for $d=4$, the time evolution of the scaled spin expectation values $X$, $Y$, $Z$ with $N\to\infty$, 
	for 20 trajectories starting at random configurations with maximal length of the total spin, 
	where the dissipation is given by $L^{(i)}_{\text{spin}}$ (a) and $L^{(i)}_{\equiv}$ (b), respectively. 
	Remember that $d=4$ is the smallest $d$ for which these two dissipators differ from each other.  
	Note furthermore that this figure is not a stream plot like Fig.~\ref{fig:qubitsMeanfield}, since the dynamics depends on further variables that are not shown.
		
	When individual dissipation is present (top and middle row of Fig.~\ref{fig:dLevelMeanfield_overview}), independently of whether there is also collective dissipation or not, the behavior is qualitatively very similar to $d=2$ (Fig.~\ref{fig:qubitsMeanfield}), for both spin-ladder dissipation 
	and $m$-independent dissipation: 
	In the dissipation-dominated regime 
	$\gamma_I + \gamma_C > 2|V|$ (right column of both subfigures), 
	all trajectories converge to the unique steady state with $X = Y = 0$, $Z = -1$. When the interaction dominates (left column of both subfigures), two steady states emerge, with $Z \neq -1$ and $X = Y$, which are identical to each other up to the sign of $X$ and $Y$. These are thus 
	broken-symmetry
	steady states like in the interaction-dominated regime for $d=2$. With increasing ratio $\gamma_C/\gamma_I$, the 
	broken-symmetry
	steady states differ further from the spin-$z$ polarized state, an effect that is slightly stronger for $m$-independent dissipation than for spin-ladder dissipation. 
	Also in comparison to $d=2$ (Fig.~\ref{fig:qubitsMeanfield}) one observes that the $d=4$ broken-symmetry steady states deviate more strongly from the spin-$z$ polarized steady state.
		
	When only collective dissipation contributes (bottom row of Fig.~\ref{fig:dLevelMeanfield_overview}), the interaction-dominated regime (bottom left panel of both subfigures) is characterized by oscillatory solutions for both types of Lindblad operators, similarly as for $d=2$.
	In fact, the behavior for spin-ladder dissipation is exactly identical to $d=2$: 
	When only collective dissipation is present and $L_C$ is a function of collective spin operators (as it is the case if $L^{(i)}$ is a linear combination of single-particle spin operators, such as $L^{(i)}_\text{spin} = j^{(i)}_{-}$), Eq.~\eqref{eq:master-eq} describes the dynamics of a single spin with conserved spin quantum number $J=Nj$ (for maximal spin length). This single spin can equivalently be constructed as the symmetric superposition of $J/j = N$ spin-$j$ particles or of $2J = 2Nj$ spin-$1/2$ particles \cite{Sakurai,CohenTannoudji} and its limit for  $N\to\infty$ is thus identical for these two scenarios.
	In contrast, $m$-independent dissipation according to $L^{(i)}_{\equiv}$ does lead to deviations from $d=2$: The total spin length is no longer conserved, but oscillates instead. This periodic change of the total spin length can be seen, e.g., from the trajectory highlighted in red in the bottom left panel of subfigure \ref{fig:dLevelMeanfield_overview}(b), which starts at the surface of the sphere, then approaches its center, before returning towards the surface again. 
		
	The fact that the total spin length is not fixed to 1 for $m$-independent collective dissipation becomes apparent also from the dynamics at dominating dissipation [bottom right panel of subfigure \ref{fig:dLevelMeanfield_overview}(b)]:
	There, multiple steady states are found with $X = Y = 0$, but different values of $Z$ depending on the initial conditions. These are the fixed points with $\left< S_{\alpha;m,n}\right> = 0$ for $m\neq n$ discussed above. 
	In contrast, for spin-ladder dissipation, only a single steady state with $X = Y = 0$, $Z = -1$ emerges, like for non-vanishing $\gamma_I$.
	
	\begin{figure}
		\includegraphics[width=\linewidth]{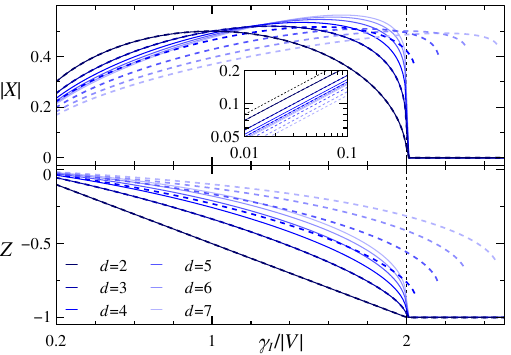}
		\caption{Scaled spin expectation values $|X| = \left<J_x\right>/Nj$ and $Z = \left<J_z\right>/Nj$ of the steady states versus $\gamma_I/|V|$ at $\gamma_C = 0$ and $N\to\infty$, for dissipators $L^{(i)}_{\text{spin}}$ (solid) and $L^{(i)}_{\equiv}$ (dashed) and for $d=2$ to $d=7$ as indicated in the legend. The black dashed line highlights the transition point $\gamma_I = 2 |V|$, below which the spin-$z$ polarized steady state (horizontal line at $X = 0$ and $Z = -1$, identical for all $d$ and both dissipators) becomes unstable. The broken-symmetry steady states for $L^{(i)}_{\equiv}$ and $d\geq 4$ remain stable above $\gamma_I = 2 |V|$ until $\gamma_I/|V| \approx 2.05$ ($d=4$), $2.16$ ($d=5$), $2.30$ ($d=6$), $2.46$ ($d=7$). The two 
		broken-symmetry
		steady states at $\gamma_I<2|V|$ 
		(and at $\gamma_I\gtrsim 2|V|$ for $L^{(i)}_{\equiv}$) 
		have identical $|X|$ and $Z$. 
		The inset shows $|X|$ for $\gamma_I/|V|\in[0.01,0.1]$ in comparison to a power law $\sim(\gamma_I/|V|)^{1/2}$ (black dashed line).
		}
		\label{fig:dLevelMeanfield-Jm_d-comparison}
	\end{figure}
		
	Despite the qualitative similarities with the qubit case, the number of single-particle levels \textit{does} have a significant effect onto the steady state, also for spin-ladder dissipation.
	This is exemplified in Fig.~\ref{fig:dLevelMeanfield-Jm_d-comparison} by the scaled steady-state spin expectation values  $|X|$ and $Z$
	for $N\to\infty$ and only individual dissipation, $\gamma_C = 0$.
	The steady states for $d>2$ are here obtained numerically by solving Eq.~\eqref{eq:mean-field} with $\left<\dot{S}_{\alpha;m,n}\right> = 0$ and then checking the stability of the resulting fixed points by inserting them into the (analytically accessible) Jacobian matrix of Eq.~\eqref{eq:mean-field}.
	For spin-ladder dissipation (solid lines), both expectation values are continuous at the transition for all $d$, whereas their slope (i.e., their first derivative with respect to $\gamma_I/|V|$) changes non-continuously at $\gamma_I = 2 |V|$.
	The transition thus remains of second order also for $d>2$.
	However, as fits with a power law $\sim(2-\gamma_I/|V|)^{\beta_{\alpha}}$ ($\alpha = X,Z$) reveal, the behavior of $X$ and $Z+1$ for $\gamma_I\lesssim 2|V|$ can be characterized by critical exponents $\beta_\alpha$ that decrease with $d$ towards 0 (i.e., for larger $d$, the spin expectation values change more rapidly at the transition). For $d=2$ to $d=7$ shown in Fig.~\ref{fig:dLevelMeanfield-Jm_d-comparison}, we find in particular $\beta_X\approx0.5$, $0.26$, $0.18$, $0.14$, $0.11$, $0.10$ and $\beta_Z\approx 1.0$, $0.55$, $0.40$, $0.33$, $0.28$, $0.25$.
	Hence, the first derivatives at $\gamma_I = 2|V|$ (except for~$Z$ with $d=2$) diverge towards infinity as $\sim (2-\gamma_I/|V|)^{-(1-\beta_\alpha)}$, but with an exponent $1-\beta_\alpha$ that increases with $d$.

	Remarkably, $m$-independent dissipation for $d\geq4$ (dashed lines) gives rise to striking differences to the qubit case and to spin-ladder dissipation for the same $d$:
	In the dissipation-dominated regime close to the transition, i.e., $\gamma_I\gtrsim 2|V|$, a bistable regime emerges in which 
	not only the spin-$z$ polarized state but also the 
	broken-symmetry
	states observed for dominating interaction (see Fig.~\ref{fig:dLevelMeanfield_overview}) are stable.
	In fact, the spin expectation values of the 
	broken-symmetry 
	steady states extend continuously from the interaction-dominated regime into the region $\gamma_I\gtrsim 2|V|$. At the upper limit of the bistable region, i.e., the point where the 
	broken-symmetry
	states finally become unstable (visible in the figure as the end points of the lines), the corresponding spin expectation values still differ from $X = 0$ and $Z = -1$, i.e., the transition is of first order, in contrast to the second-order transition observed for $d=2$ and for spin-ladder dissipation.
	As $d$ increases, the bistable region grows (from $\gamma_I/|V|\in [2,2.05]$ for $d=4$ to $\gamma_I/|V|\in [2,2.46]$ for $d=7$), and the spin expectation values of the 
	broken-symmetry
	states at the upper limit of the bistable region differ further from those of the spin-$z$ polarized state.
	Note that a bistable region has recently been found also for the Dicke model with individual and collective dissipation \cite{Leppenen2024}. In that model, however, bistability emerges already for $d=2$, in contrast to our findings reported here.
	
	Far from the transition point, the absolute value of the spin expectation values decreases as $\gamma_I$ is reduced, for both $L^{(i)}_{\text{spin}}$ and $L^{(i)}_{\equiv}$. 
	Comparison of the data for $\gamma_I\gtrsim 0$ to a power law (exemplified for $X$ in the inset of Fig.~\ref{fig:dLevelMeanfield-Jm_d-comparison}) reveals that $X \sim (\gamma_I/|V|)^{1/2}$, $Z \sim \gamma_I/|V|$ to leading order in $\gamma_I/|V|$, for both dissipators and all $d$, and consequently $\lim_{\gamma_I\to 0} X = \lim_{\gamma_I\to 0} Z = 0$ like for $d=2$.
	However, the prefactor of this convergence 
	decreases as $d$ is increased, for $X$ from $1/\sqrt{2}$ ($d=2$) to 0.472 ($d=7$, spin-ladder dissipation) and 0.378 ($d=7$, $m$-independent dissipation), respectively, and for $Z$ from $-1/2$ ($d=2$) to $-0.087$ ($d=7$, spin-ladder dissipation) and $-0.073$ ($d=7$, $m$-independent dissipation), respectively.
		
	Note that due to the symmetries of the system any result discussed in this section for $\left|X\right|$ is valid also for $|Y| = |X|$, and the findings about the 
	broken-symmetry
	states hold for both of these states, since they differ from each other only in the sign of $X$ and $Y$. 
	
	We can finally conclude for spin-ladder dissipation that the properties of the steady state for $d>2$ remain qualitatively the same as for $d=2$, 
	while features such as the divergence from the spin-$z$ polarized state for $\gamma_I\lesssim 2|V|$ and the convergence to $r=0$ for $\gamma_I \to 0$ become more prominent with increasing $d$. 
	For $m$-independent dissipation, in contrast, the phase diagram changes also qualitatively through the emergence of the bistable region.

	\subsection{Large-Spin Limit for Spin-Ladder Dissipation}
	
	As we have seen in 
	the \hyperref[sec:Qudits]{preceding section} 
	for $d=2$ to $d=7$, the features of the steady state for spin-ladder dissipation remain qualitatively unchanged as $d$ increases in the limit $N\to\infty$.
	Motivated by this result, we will now investigate whether these findings are still valid for an infinite number of single-particle levels and whether, consequently, the two limits $N\to\infty$ and $d\to\infty$ commute.
	To answer these questions, we consider the limit $j\to\infty$ of scaled spins 
	$j_\alpha^{(i)}/{j}$ ($i=1,\ldots,N$, $\alpha = x,y,z$),
	where $N$ is finite. Such a limit can, for instance, be achieved by $N$ collective spins of a large number of atoms, in a similar fashion as, e.g., the models discussed in Refs.~\cite{Huber2020,SilvaSouza2023}. 
	
	One can easily check that $\langle j_\alpha^{(i)}\rangle/j = \O(1)$, whereas $\langle [j_\alpha^{(i)}/j,j_\beta^{(k)}/j]\rangle = \i \delta_{ik} \sum_{\gamma=x,y,z} \varepsilon_{\alpha\beta\gamma} \langle j_\gamma^{(i)}\rangle/j^2$ (with $i,k=1,\ldots,N$ and $\alpha,\beta = x,y,z$),
	converges to 0 as $j\to\infty$. 
	In this limit, we can thus 
	apply a mean-field assumption 
	in the equations of motion for the scaled spins, which, with $\alpha^{(i)} = \langle j_\alpha^{(i)} \rangle/{j}$ ($i=1,\ldots,N$, $\alpha = x,y,z$), leads to 
	\begin{subequations}
	\begin{align}
		\dot{x}^{(i)} ={}& - \frac{2 V}{N}\left(\sum_{k=1}^N y^{(k)}\right) z^{(i)} + \gamma_I x^{(i)} z^{(i)} + \frac{\gamma_C}{N} \left(\sum_{k=1}^N x^{(k)}\right) z^{(i)},\\
		\dot{y}^{(i)} ={}& -\frac{2V}{N} \left(\sum_{k=1}^N x^{(k)}\right) z^{(i)} + \gamma_I y^{(i)} z^{(i)} + \frac{\gamma_C}{N} \left(\sum_{k=1}^N y^{(k)}\right) z^{(i)},\\
		\nonumber
		\dot{z}^{(i)} ={}& \frac{2V}{N} \sum_{k=1}^N\left( x^{(k)} y^{(i)} + y^{(k)} x^{(i)} \right) - \gamma_I\left[\left(x^{(i)}\right)^2 + \left(y^{(i)}\right)^2 \right] \\
		&- \frac{\gamma_C}{N}\sum_{k=1}^N\left(x^{(k)} x^{(i)} + y^{(k)}y^{(i)}\right).
	\end{align}
	\end{subequations}
	While the squared length $\left(x^{(i)}\right)^2 + \left(y^{(i)}\right)^2 + \left(z^{(i)}\right)^2 = (j+1)/j$ of the single-particle spins is constant throughout time evolution, the squared length $\left(\sum_{i=1}^N x^{(i)}\right)^2 + \left(\sum_{i=1}^N y^{(i)}\right)^2 + \left(\sum_{i=1}^N z^{(i)}\right)^2$ of the total spin is not necessarily conserved. However, if the system is initialized in a state with maximum total spin length, then $\alpha^{(i)}=\alpha^{(k)}$ ($i,k=1,\ldots,N$, $\alpha = x,y,z$) for all times~\footnote{At initial time, this condition can be shown, e.g., by the Cauchy-Schwarz inequality. Assuming this condition for all times leads to a valid solution of the set of differential equations. According to the Picard-Lindelöf theorem, this solution is the unique solution.}.
	Then
	\begin{subequations}
	\begin{align}
		\dot{x}^{(i)} ={}& - 2V y^{(i)} z^{(i)} + \left(\gamma_I+\gamma_C\right) x^{(i)} z^{(i)},\\
		\dot{y}^{(i)} ={}& - 2V x^{(i)} z^{(i)} + \left(\gamma_I+\gamma_C\right) y^{(i)} z^{(i)},\\
		\dot{z}^{(i)} ={}& 4V x^{(i)} y^{(i)} - \left(\gamma_I+\gamma_C\right)\left[\left(x^{(i)}\right)^2 + \left(y^{(i)}\right)^2 \right],
	\end{align}
	\end{subequations}
	which is nothing else than the mean-field equations for $N\to\infty$ and $d=2$, Eq.~\eqref{eq:qubits},
	with collective dissipation at the rate $\gamma_I + \gamma_C$ and vanishing individual dissipation. 
	Hence, the transition point remains at $\gamma_I + \gamma_C = 2|V|$ like for $N\to\infty$ at finite $d$, but the transition is now of first order also for finite $\gamma_I$ and the steady states in the interaction-dominated regime are always oscillatory, no matter whether $\gamma_I$ is present or not. 
	Consequently the steady-state properties for a finite number of infinitely long collective spins (i.e., $d\to\infty$ at fixed $N$) are strikingly different from those for an infinite number of particles with finite spin (i.e., $N\to\infty$ at fixed $d$), even though the model for finite $N$ and $d$ is exactly the same.

\section{Spectral Fingerprints of the Dissipative Phase Transition}

\label{sec:numericsL}

In the following sections, we investigate numerically how the different phases manifest themselves at finite particle numbers $N$, restricting to $\gamma_I \neq 0$ for simplicity. 
To this end, we interpret the right-hand side of Eq.~\eqref{eq:master-eq} as 
a linear superoperator $\LL$ (the Liouvillian) acting on the density matrix $\rho$.
The steady states are then eigenmatrices of $\LL$ with eigenvalue~0.
We express the Liouvillian as a $(D\times D)$-dimensional matrix and employ 
its permutation invariance
\cite{Gegg2016,GeggThesis,HuybrechtsThesis,Sukharnikov2023} to reduce its matrix dimension from $D = d^{2N}$ to $\binom{N+d^2-1}{N}$, 
which scales with $N$ as $N^{d^2 - 1}$, i.e., polynomially.
Due to the $\Z_2$ symmetries $\Pi_1$ and $\Pi_2$, $\LL$ is block-diagonal with four blocks of approximately equal size, 
i.e., of dimensions that are further reduced by a factor $\approx 1/4$ compared to the full Liouvillian dimension. 
Nevertheless, since the order of this polynomial grows quadratically with $d$, the numerics remain limited to rather small particle numbers already for moderately large~$d$: 
For instance, blocks of dimension $\approx\num{150000}$ correspond to $N\approx 150$ two-level systems, but to only $N=6$ five-level systems [exact values of the four block dimensions: $\num{149226}$, $\num{143450}$, $\num{146300}$, $\num{146300}$ for $(N,d) = (150,2)$ and $\num{149535}$, $\num{147580}$, $\num{148330}$, $\num{148330}$ for $(N,d) = (6,5)$].

From the symmetry-resolved (with respect to $\Pi_1$ and $\Pi_2$) spectrum of $\LL$, one can detect dissipative phase transitions and symmetry breaking \cite{Minganti2018,Minganti2021}: 
Within each symmetry sector induced by the $\Z_2$ symmetries, here labelled by parities $p_k\in\{+1,-1\}$ with respect to $\Pi_k$ $(k=1,2)$,
let $\lambda_i^{(p_1,p_2)}$ be the eigenvalues of $\LL$, sorted such that $|\Re[\lambda_i^{(p_1,p_2)}]|\leq |\Re[\lambda_{i+1}^{(p_1,p_2)}]|$ ($i = 0,1,2,\ldots$). 
The steady state is contained in the fully symmetric sector with respect to the $\Z_2$ symmetries \cite{Minganti2018}, 
which is here the space characterized by  parities $p_1 = p_2 = +1$. 
When a symmetry of the equations of motion is broken by the steady states, the gaps
\begin{align}
	\Delta^{(p_1,p_2)} = -\Re\big[\lambda_0^{(p_1,p_2)}\big], \quad (p_1,p_2) \neq (+1,+1)
\end{align}
between the eigenvalue $\lambda_0^{(+1,+1)} = 0$ of the steady state and the lowest eigenvalue $\lambda_0^{(p_1,p_2)}$ in the sector $(p_1,p_2)$ vanish as $N\to\infty$, for all sectors $(p_1,p_2)$ corresponding to the broken symmetry \cite{Minganti2018,Minganti2021}.
In our model, the steady states remain symmetric with respect to $\Pi_2$ for all $V>0$  and with respect to $\Pi_1\Pi_2$ for all $V<0$, while the symmetry $\Pi_1$ is spontaneously broken in the thermodynamic limit for $\gamma_I + \gamma_C < 2 |V|$, see Section~\ref{sec:mean-field}. The symmetry sector corresponding to the symmetry breaking at $V>0$ ($V<0$) is hence the one, where the parity of $\Pi_1$ differs from the fully symmetric space and the parity of $\Pi_2$ (of $\Pi_1\Pi_2$) remains the same, i.e., $(p_1,p_2) = (-1,+1)$ [$(p_1,p_2) = (-1,-1)$].
Furthermore, 
when several steady states not related by symmetry coexist in the thermodynamic limit for the same parameter values, e.g., at the critical point of a first-order phase transition \cite{Minganti2018}, also the gap 
\begin{align}
	\Delta^{(+1,+1)} = -\Re[\lambda_1^{(+1,+1)}]
\end{align}
between $\lambda_0^{(+1,+1)} = 0$ and the first nonzero eigenvalue $\lambda_1^{(+1,+1)}$ of the same symmetry sector vanishes as $N\to\infty$. 

\begin{figure}
	\includegraphics{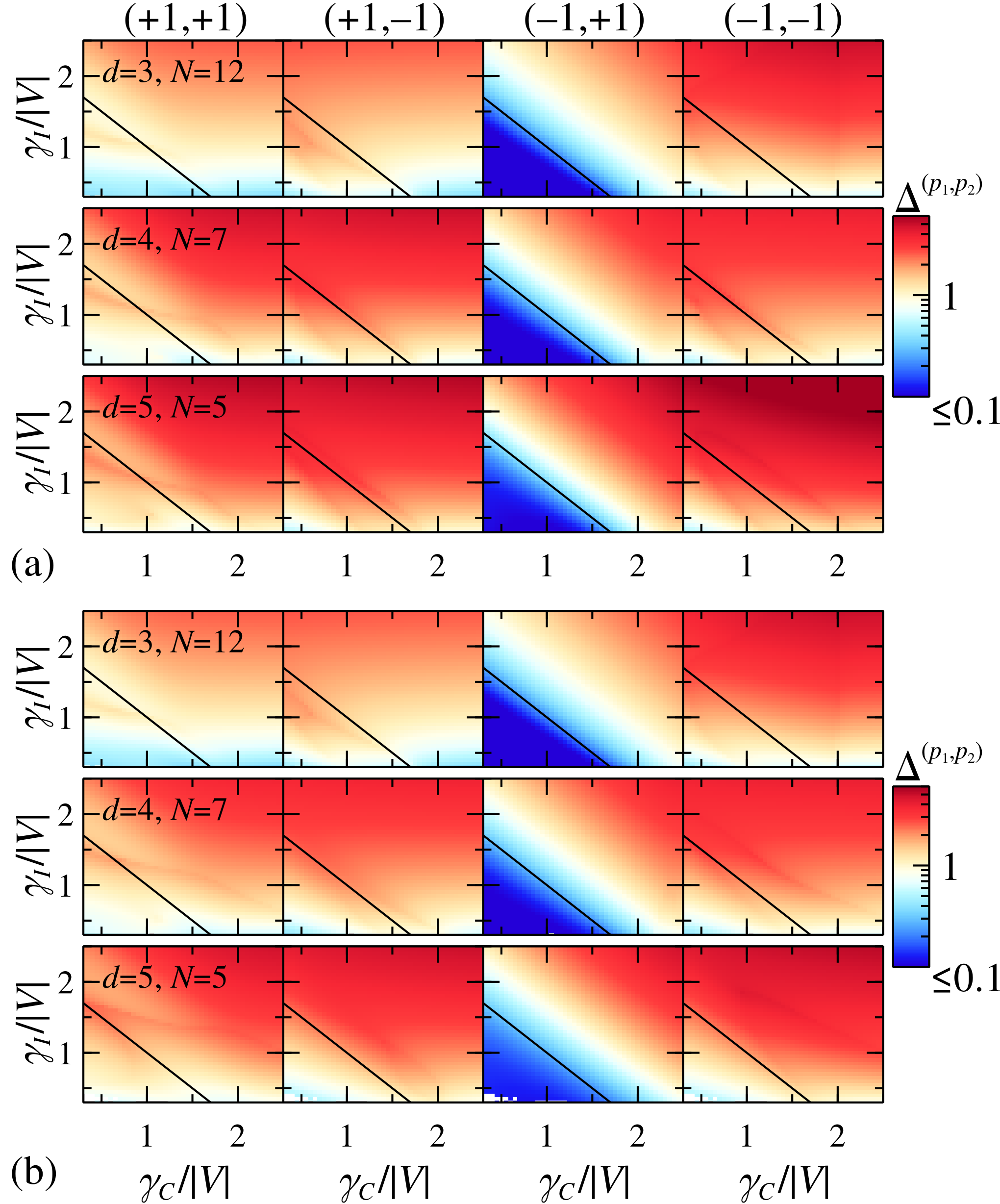}
	\caption{Gaps $\Delta^{(p_1,p_2)}$ for spin-ladder (a) and $m$-independent (b) dissipation, versus $\gamma_I/|V|$ and $\gamma_C/|V|$ (only $V\geq 0$), with parities $(p_1,p_2)$ as indicated in the top row and for different $N$, $d$ (rows) such that the dimension of the Liouvillian is comparable. The color bar is identical for all panels. The black line indicates the mean-field transition point $\gamma_I + \gamma_C = 2|V|$. White areas denote parameter values, for which not all the necessary eigenvalues could be computed. Since $L^{(i)}_\text{spin} = L^{(i)}_\equiv$ for $d=3$, the upper rows of subfigures (a) and (b) are identical.
	}
	\label{fig:GapClosing1}
\end{figure}

Figure~\ref{fig:GapClosing1} shows the values of the gaps $\Delta^{(p_1,p_2)}$ for all symmetry sectors, both dissipators $L^{(i)}_{\text{spin}}$, $L^{(i)}_{\equiv}$, and $d=3$ to $d=5$, with particle numbers $N$ such that the dimensions of the symmetry sectors 
are between $\approx \num{30000}$ and $\approx \num{40000}$ for all~$d$. 
Even for the small particle numbers considered here, the gap $\Delta^{(-1,+1)}$ clearly closes in the interaction-dominated phase where symmetry breaking occurs in the thermodynamic limit, for both types of dissipators. Furthermore, for $m$-independent dissipation and $d = 4,5$, the region where $\Delta^{(-1,+1)}$ significantly decreases appears to be slightly shifted towards larger $\gamma_I+\gamma_C$ as compared to spin-ladder dissipation, most prominently visible for $d=N=5$. 
This suggests that $\Delta^{(-1,+1)} \to 0$ $(N\to \infty)$ also in the bistable region, 
as expected from the existence of 
broken-symmetry
steady states in that domain.

In contrast, the behavior of the other gaps $\Delta^{(+1,+1)}$, $\Delta^{(\pm 1,- 1)}$ is less clear, and a significant closing of the gaps in the interaction-dominated phase is not observed. Nevertheless, $\Delta^{(+1,+1)}$ and $\Delta^{(\pm 1,- 1)}$ are reduced 
as compared to the dissipation-dominated phase.

Note that there is a tendency of all gaps to close for \mbox{$\gamma_I, \gamma_C\to 0$}. This can be understood from the exact limit $\gamma_I=\gamma_C = 0$: There, only the unitary dynamics survives, which in the thermodynamic limit has four steady states related to each other by symmetry [given by Eq.~\eqref{eq:steady-state-qubits-collective} for $\gamma_C = 0$]. Consequently, both symmetries $\Pi_1$, $\Pi_2$ are broken and we expect all gaps $\Delta^{(p_1,p_2)}$ with $(p_1,p_2)\neq (+1,+1)$ to close.

\begin{figure*}
	\includegraphics{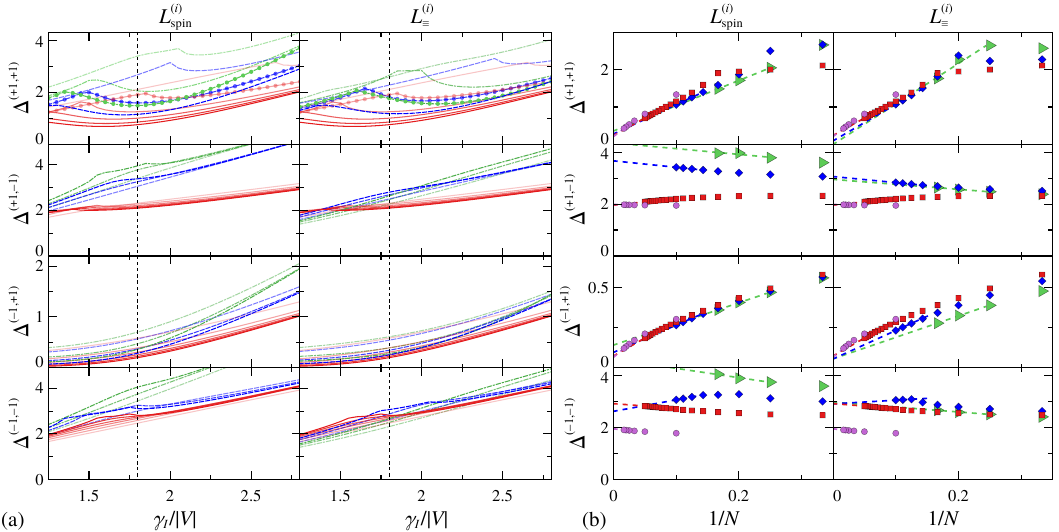}
	\caption{Gaps $\Delta^{(p_1,p_2)}$ at $\gamma_C/|V|=0.2$, for spin-ladder and $m$-independent dissipation as indicated in the top row: (a) versus $\gamma_I/|V|$, for $d=3$, $N=3,6,9,12,15,18$ (red solid), $d=4$, $N=3,6,9$ (blue dashed), $d=5$, $N=2,4,6$ (green dash-dotted), where stronger color means larger particle number and $N=6$ is indicated for $\Delta^{(+1,+1)}$ by circles accompanying the lines, (b) versus $1/N$ at the transition point $\gamma_I/|V|=1.8$
	[black dashed line in (a)], for $d=2$ (purple circles), $d=3$ (red squares), $d=4$ (blue diamonds), $d=5$ (green triangles). Dashed lines in (b) are linear fits to the last three points. Since $L^{(i)}_\text{spin} = L^{(i)}_\equiv$ for $d=2,3$, the red lines in (a) and the red and purple data points in (b) are identical for both dissipators.}
	\label{fig:GapClosing2}
\end{figure*}

To get a clearer understanding of the gaps also for $(p_1,p_2)\neq(-1,+1)$, we plot $\Delta^{(p_1,p_2)}$ in Fig.~\ref{fig:GapClosing2} as functions of $\gamma_I/|V|$ and of particle number at constant $\gamma_C/|V| = 0.2$ (i.e., along the vertical direction of Fig.~\ref{fig:GapClosing1} for dominating individual dissipation), for spin-ladder and $m$-independent dissipation. 
All four gaps are found to decrease as $\gamma_I/|V|$ is reduced. However, no trace of the phase transition is visible in the two gaps $\Delta^{(\pm 1,- 1)}$, except for a small kink in the vicinity of the critical point $\gamma_I/|V|=1.8$ for some $(N,d)$. But note that for other $(N,d)$ such a feature is either absent or appears far from the critical point, i.e., this kink might as well be unrelated to the phase transition and due to finite-size effects. Furthermore, increasing the number of particles at constant $d$ does not significantly reduce these gaps in the interaction-dominated phase. In fact, as visible from Fig.~\ref{fig:GapClosing2}(b) directly at the critical point $(\gamma_I/|V|,\gamma_C/|V|) = (1.8,0.2)$, the gaps $\Delta^{(\pm 1,- 1)}$ may even slightly increase with particle number, for both types of dissipation.
These results clearly show that the symmetry $\Pi_2$ remains intact at the phase transition.

In contrast, $\Delta^{(+1,+1)}$ and $\Delta^{(-1,+1)}$ do display features of the phase transition: Similarly to Fig.~\ref{fig:GapClosing1}, the symmetry-related gap $\Delta^{(-1,+1)}$ is greatly reduced in the interaction-dominated phase and decreases strongly with increasing particle number. Note, however, that an extrapolation to infinite $N$ at $(\gamma_I/|V|,\gamma_C/|V|) = (1.8,0.2)$ [dashed lines in Fig.~\ref{fig:GapClosing2}(b), obtained as linear fits to the three largest $N$] suggests small but still finite gaps $\Delta^{(-1,+1)}\in[0.066,0.142]$ for spin-ladder dissipation and $\Delta^{(-1,+1)}\in[0.056,0.076]$ for $m$-independent dissipation. This is probably a finite-size effect due to the limited range of numerically accessible particle numbers.

As a function of $\gamma_I/|V|$, the gap $\Delta^{(+1,+1)}$ within the fully symmetric sector of the $\Z_2$ symmetries shows a pronounced minimum, which for sufficiently large $N$ is found directly at the transition for spin-ladder dissipation and at slightly larger $\gamma_I/|V|$ for $m$-independent dissipation and $d=4,5$. 
With increasing $N$, the gap keeps reducing for all $\gamma_I$ investigated, and an extrapolation from the available particle numbers to $N\to\infty$ at $(\gamma_I/|V|,\gamma_C/|V|) = (1.8,0.2)$ yields gaps $\Delta^{(+1,+1)}\in[0.207,0.357]$ for spin-ladder dissipation and $\Delta^{(+1,+1)}\in[-0.006,0.237]$ for $m$-independent dissipation. 
The data thus suggests that the gap within the same symmetry sector might close as $N\to\infty$, at least at the critical point, even though this cannot clearly be confirmed from the available finite-size data. 
Note that, for $d=4,5$ and $m$-independent dissipation, the minimal $\Delta^{(+1,+1)}$ is at slightly larger $\gamma_I/|V|$ than the critical point $\gamma_I/|V| = 1.8$.
Hence, if the gap closes directly at the critical point, then $\Delta^{(+1,+1)} \to 0$ ($N\to\infty$) also in a region directly above $\gamma_I/|V| = 1.8$.

As discussed at the beginning of the section, we expect such closing of the gap $\Delta^{(+1,+1)}$ if several steady states not related by symmetry are present in the thermodynamic limit. The latter is the case in the bistable region for $m$-independent dissipation and $d\geq 4$, where the 
broken-symmetry
and the spin-$z$ polarized steady states coexist in the thermodynamic limit and where our data indeed suggests $\Delta^{(+1,+1)}\to 0$ ($N\to\infty$). Interestingly, our data is compatible with a closing of the gap directly at the transition $\gamma_I + \gamma_C = 2|V|$ for both types of dissipation. If $\Delta^{(+1,+1)}$ indeed vanishes there, this means that the symmetry breaking is not the only mechanism inducing the second-order phase transition in the LMG model for spin-ladder dissipation, but instead it is accompanied by a non-analytic change of the steady state within the symmetric sector of the $\Z_2$ symmetries. This finding is in line with Ref.~\cite{Minganti2021}, where it is shown that symmetry breaking can be removed from a second-order dissipative phase transition without removing the transition itself.

When comparing different $d$ in the regions where the gaps $\Delta^{(+1,+1)}$ and $\Delta^{(-1,+1)}$ close as $N\to\infty$, one finds that this convergence to 0 is faster with increasing $d$.
This is exemplified by $\Delta^{(+1,+1)}$ for $N=6$ and $d=3,4,5$ [indicated in Fig.~\ref{fig:GapClosing2}(a) by the filled circles accompanying the lines], where the minimum of the gap as a function of $\gamma_I/|V|$ is deeper for larger $d$, and it is also visible from Fig.~\ref{fig:GapClosing2}(b), where $\Delta^{(+1,+1)}$ and $\Delta^{(-1,+1)}$ tend to decrease with $d$ once $N$ is large enough.
We can explain this finding from the fact that the dimension of $\LL$ grows with $d$ at fixed particle numbers, which means that for larger $d$ the system is effectively closer to the thermodynamic limit (where we expect the gaps to close exactly).

The spectrum thus reveals the symmetry breaking in the interaction-dominated phase and the bistable region, and hints towards the coexistence of steady states not related by symmetry (in the bistable region for $m$-independent dissipation) and towards a non-analytic change of the steady state at the second-order transition (for spin-ladder dissipation). A larger number $d$ of single-particle levels makes the corresponding spectral features more prominent for finite particle numbers.

\section{Purity and Entanglement in the Steady State}

\label{sec:numericsRho}

To further understand the properties of the different phases at finite $N$, we numerically investigate also the steady states themselves, focusing in particular on their purity and on their bipartite entanglement properties, which we here quantify by the negativity \cite{Zyczkowski1998,Vidal2002,Horodecki2009,Guhne2009}.
Note that for finite $N$ and $\gamma_I\neq 0$ our model always hosts a unique steady state $\rho_{\text{ss}}$~\cite{Nigro2019} (provided that all $\ell_m\neq 0$), 
despite the existence of several steady states for the interaction-dominated and bistable phases in the limit $N\to\infty$. 
The finite-size steady state therefore has to be a mixture of the mean-field steady states.

\begin{figure}
	\centering
	\includegraphics{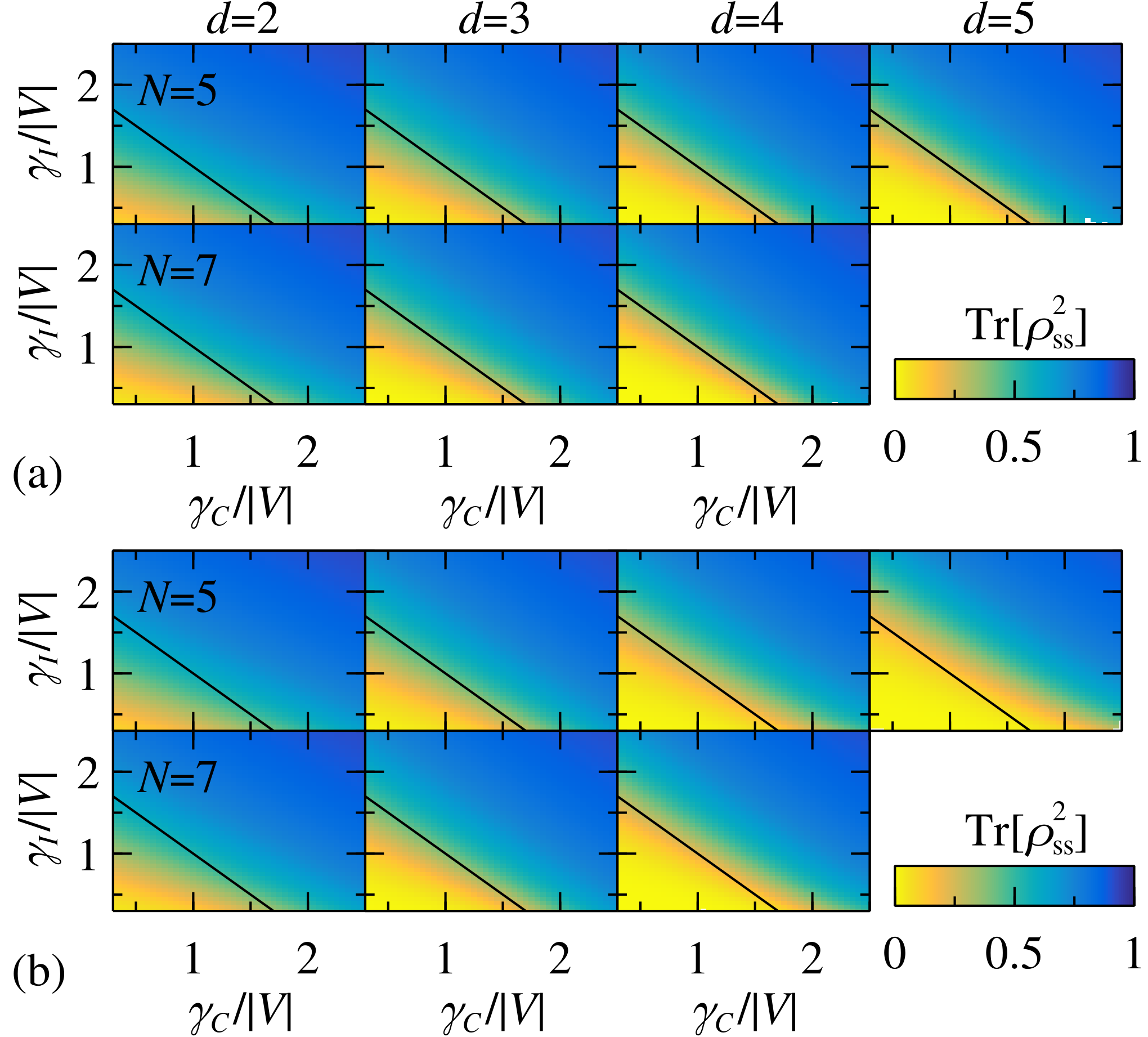}
	\caption{Steady-state purity $\Tr[\rho_{\text{ss}}^2]$ for spin-ladder (a) and $m$-independent (b) dissipation, versus $\gamma_{C}/|V|$ and $\gamma_{I}/|V|$, for particle numbers $N=5$ and $N=7$ (rows) and single-particle dimensions $d=2$ to $d=5$ (columns). 
	The color bar is identical for all panels. The black line indicates the mean-field critical point $\gamma_I + \gamma_C = 2|V|$. 
	Since $L^{(i)}_\text{spin} = L^{(i)}_\equiv$ for $d=2,3$, the two leftmost rows of subfigures (a) and (b) are identical. 
	Liouvillian dimensions [sector $(p_1,p_2)=(+1,+1)$] range up to \num{30087} ($N=d=5$) and \num{43186} ($N=7, d=4$).}
	\label{fig:Purities}
\end{figure}

Figure~\ref{fig:Purities} shows the purity $\Tr[\rho_{\text{ss}}^2]$ of the numerically computed steady state $\rho_{\text{ss}}$ as a function of $\gamma_{C}/|V|$ and $\gamma_{I}/|V|$, for $d=2$ to $d=5$ and $N=5$, $N=7$. As $\gamma_{C}/|V|$ or $\gamma_{I}/|V|$ is increased, the steady state turns from a highly mixed state with $\Tr[\rho_{\text{ss}}^2]\ll 1$ into an almost pure state with $\Tr[\rho_{\text{ss}}^2]\approx 1$, with a drastic change directly at the phase transition. This rise of the purity at the transition gets ever sharper as $N$ or $d$ are increased.
While for spin-ladder dissipation [subfigure (a)] the boundary between highly mixed and approximately pure steady states is almost exactly at $\gamma_I + \gamma_C = 2|V|$, it is slightly shifted to larger $\gamma_I + \gamma_C$ for $d=4, 5$ and $m$-independent dissipation [subfigure (b)] and this shift increases from $d=4$ to $d=5$. Consequently, for $m$-independent dissipation the switch from a highly mixed to an almost pure steady state is related to the transition between the bistable region and the dissipation-dominated phase. 

The observed behavior of the purity is in line with the mean-field results: 
Since the unique steady state $\rho_{\text{ss}}$ for finite $N$ is a mixture of all the mean-field steady states,
it is a mixed state (and consequently $\Tr[\rho_{\text{ss}}^2]<1$) once two or more steady states exist in the thermodynamic limit. As discussed in Sec.~\ref{sec:mean-field}, this is the case exactly in the interaction-dominated and bistable phases.
For the dissipation-dominated phase, the pure state $\rho = \left(\ket{-j}\bra{-j}\right)^{\otimes N}$ of $N$ particles in level $m=-j$ can be shown to be the steady state in the exact limit $V=0$ and we can thus expect the steady state to be close to this state (and hence to be almost pure) also for small but finite $V$.

Note that the mean-field assumptions of Sec.~\ref{sec:mean-field} imply that the steady states for $N\to\infty$ are well approximated by product states $\varrho^{\otimes N}$ of identical single-particle density matrices $\varrho$ for all particles and the finite-$N$ steady state for $N\gg 1$ is thus approximately of the form $\sum_{i=1}^n p_i \varrho_i^{\otimes N}$ with $n$ the number of mean-field steady states. Hence, its purity is approximately $\sum_{i,k=1}^n p_i p_k \Tr[\varrho_i \varrho_k]^{N}$ and thus decays exponentially with $N$ once $\Tr[\varrho_i \varrho_k]<1$ for all $i,k$. This explains the enhanced contrast between the dissipation-dominated and the other two phases as $N$ is increased.

For the sharpening with $d$, note that $\Tr[\rho_{\text{ss}}^2]\geq 1/\operatorname{rank}\rho_{\text{ss}}$, where $\operatorname{rank}\rho_{\text{ss}}\leq d^N$ is the rank of $\rho_{\text{ss}}$. 
The rank thus yields a rough estimate of the purity, which, however, is exact only if $1/\operatorname{rank}\rho_{\text{ss}}$ is the only nonzero eigenvalue of $\rho_{\text{ss}}$.
Since any diagonal state in the eigenbasis of $H$ constitutes a steady state in the exact limit $\gamma_I = \gamma_C = 0$, we can expect the steady state at small but finite dissipation to be close to such a diagonal state over a large number of eigenstates and to thus have almost maximal rank, $\operatorname{rank}\rho_{\text{ss}} = d^N$, such that its purity decreases with $d$ roughly as $\Tr[\rho_{\text{ss}}^2]\sim d^{-N}$. 
Indeed, for $N$ and $d$ as in Fig.~\ref{fig:Purities} and $\gamma_I = \gamma_C = 0.35 |V|$ (lower left corner of the panels) we find $d^N \Tr[\rho_{\text{ss}}^2] \in [3.41,14.22]$, i.e., the purity deep in the interaction-dominated phase is approximately of the order~$d^{-N}$.

\begin{figure}
	\includegraphics{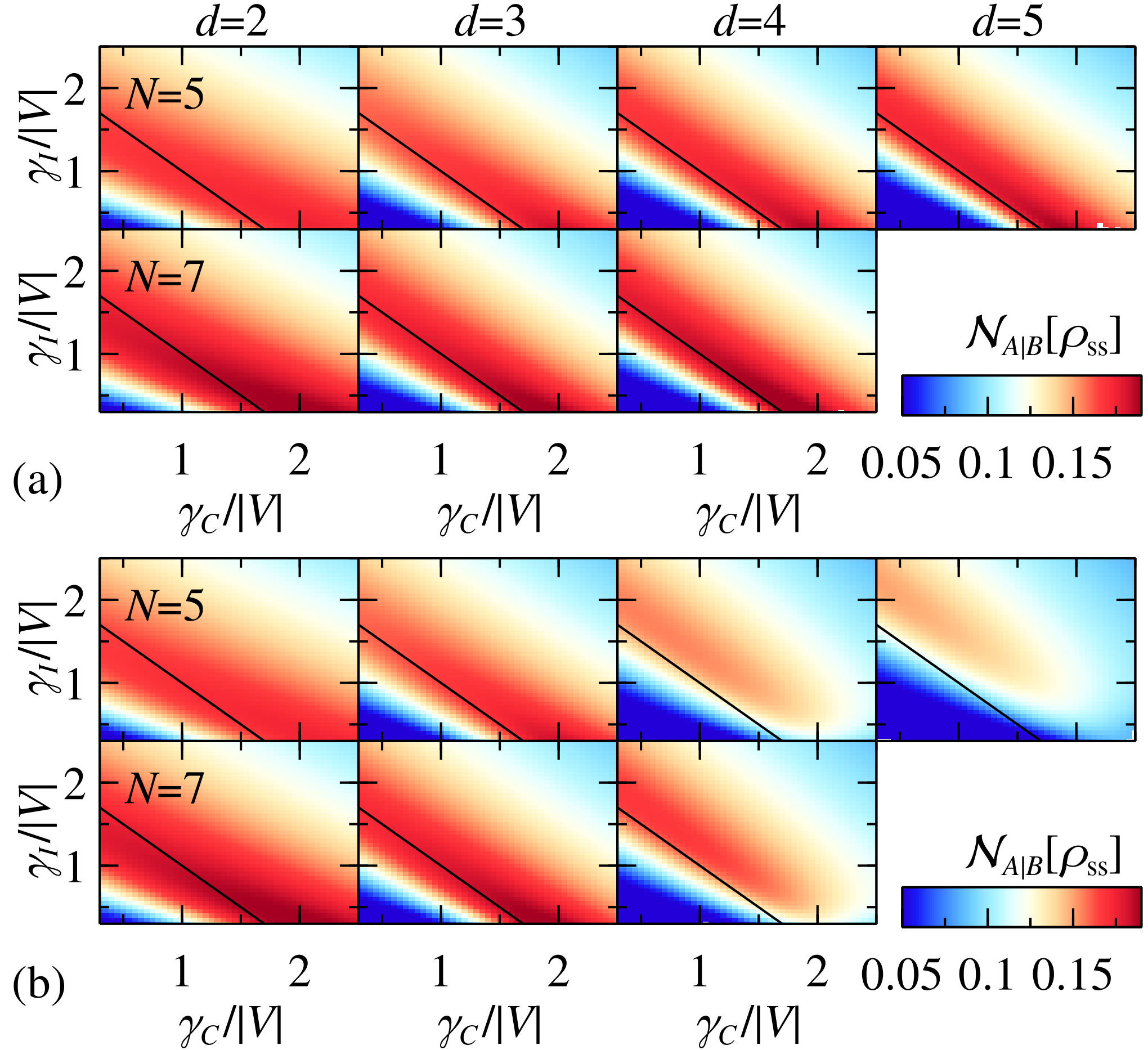}
	\caption{Steady-state negativity with respect to a bipartition into $N_A = (N+1)/2$ and $N_B = (N-1)/2$ particles, for spin-ladder (a) and $m$-independent (b) dissipation, versus $\gamma_{C}/|V|$ and $\gamma_{I}/|V|$, for particle numbers $N=5$ and $N=7$ (rows) and single-particle dimensions $d=2$ to $d=5$ (columns). The color bar is identical for all panels. The black line indicates the mean-field critical point $\gamma_I + \gamma_C = 2|V|$.
	Since $L^{(i)}_\text{spin} = L^{(i)}_\equiv$ for $d=2,3$, the two leftmost rows of subfigures (a) and (b) are identical. 
	Liouvillian dimensions [sector $(p_1,p_2)=(+1,+1)$] range up to \num{30087} ($N=d=5$) and \num{43186} ($N=7, d=4$).}	
	\label{fig:negativity}
\end{figure}

To further reveal the impact of the dissipative phase transition on the steady state, we investigate its entanglement negativity $\mathcal{N}_{A|B}[\rho_{\text{ss}}]$ 
with respect to bipartitions into subsystems $A$ and $B$ of $N_A = \left\lceil\frac{N}{2}\right\rceil$ and $N_B = \left\lfloor \frac{N}{2}\right\rfloor$ particles. $\mathcal{N}_{A|B}[\rho]$ is defined as \cite{Zyczkowski1998,Vidal2002,Horodecki2009,Guhne2009}
\begin{align}
	\mathcal{N}_{A|B}[\rho] = \sum_{\substack{\lambda \text{ eigenvalue of }\rho^{T_B},\\ \lambda < 0}} |\lambda|,
\end{align}
where $\rho^{T_B}$ is the partial transpose of $\rho$ with respect to subsystem~$B$. 
Note that, due to permutation invariance, the bipartite entanglement properties of $\rho_{\text{ss}}$ are exactly the same for all bipartitions with the same particle numbers $N_A$, $N_B$, i.e., there is no need to distinguish them.

Figure~\ref{fig:negativity} shows $\mathcal{N}_{A|B}[\rho]$ for the steady state as a function of $\gamma_{C}/|V|$ and $\gamma_I/|V|$ for $N=5$, $N=7$ and $d=2$ to $d=5$, i.e., the same $N$ and $d$ as in Fig.~\ref{fig:Purities}. For both dissipators and for all $N$ and $d$ investigated here, the negativity shows a pronounced maximum on the dissipation-dominated side of the phase transition, which slightly increases as a function of $N$. As a function of $d$, the negativity maximum gets sharper---most prominently from $d=2$ to $d=3$---whereas its height is rather robust against changes of $d$. 
Independently of the ratio of $\gamma_I$ to $\gamma_C$, the maximal negativity for spin-ladder dissipation is always larger than the corresponding value for $m$-independent dissipation.
If one aims to prepare as much steady-state entanglement as possible, the former type of dissipation should therefore be preferred over the latter.

While such an entanglement maximum at the phase transition thus emerges for both dissipators 
and has in fact been found 
also in other models \cite{Schneider2002,GonzalezTudela2013,Lee2013a,Lee2014a,Wolfe2014,Barberena2019,Wang2023e}, 
its dependence
on $\gamma_I$ and $\gamma_C$ is strikingly different for the two dissipators: With spin-ladder dissipation [subfigure (a)] the largest negativity at the phase transition is found for dominating collective dissipation, while the presence of individual dissipation slightly reduces the maximal negativity. In contrast, for $m$-independent dissipation and $d\geq 4$ [subfigure (b)], the negativity maximum is largest when $\gamma_I$ dominates and collective dissipation actually leads to a strong suppression of negativity. 
This is a surprising result, given that individual dissipation on its own,
due to its local nature, cannot induce any entanglement, but instead breaks entanglement after sufficiently long times~\cite{Khatri2020}.
(Note, however, the model of Ref.~\cite{Tucker2020}, where individual dissipation enhances entanglement on transient time scales.)

\begin{figure}
	\includegraphics{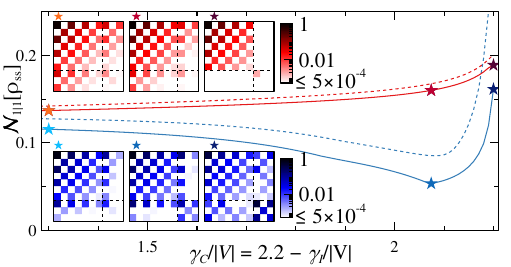}
	\caption{Steady-state negativity for $(N,d)=(2,4)$ and a bipartition into $N_A=N_B=1$, along the line $\gamma_C+\gamma_I=2.2$, for spin-ladder (red) and $m$-independent (blue) dissipation. Solid lines correspond to full-system negativities and dashed lines denote the negativity of the restriction $\rho_{\text{ss,sym}}$ to the symmetric subspace. Insets show $|\rho_{\text{ss,sym}}|$ in the basis $\ket{J,M}$, $J=3,1$, $M=-J,\ldots,J$ (sorted from $\ket{J,-J}$ to $\ket{J,J}$), for specific $(\gamma_C,\gamma_I)$ marked by stars of the same color as depicted above each matrix. Dashed lines in the insets separate $J=3$ and $J=1$.}
	\label{fig:NegativityN2d4}
\end{figure}

A first understanding of this surprising result might be obtained from comparing the effects of unitary evolution and collective dissipation on the density matrix to those of individual dissipation: Whereas $H$ and $L_C$ are block-diagonal with respect to the subspaces induced by permutation symmetry (such as the symmetric subspace given by $\pi\rho=\rho\pi=\rho$ for all permutations $\pi$) and therefore do not mix the dynamics of different subspaces, individual dissipation may induce transitions between these subspaces. In comparison to collective $m$-independent dissipation, the dynamics induced by collective spin-ladder dissipation is even more restricted, as $J_-$ is block-diagonal even with respect to the (smaller) subspaces spanned by collective spin states $\ket{J,M,\alpha}$, $M=-J,\ldots,J$, where $\alpha$ labels different subspaces characterized by the same collective spin quantum number $J$. 
To check the relation between the negativity and the different structures of collective and individual dissipation,
we study 
the simplest system for which we expect 
a behavior of negativity like for $d=4,5$ in Fig~\ref{fig:negativity},
i.e., $N=2$ particles with $d=4$ levels,
and we restrict to the specific line $(\gamma_C+\gamma_I)/|V|=2.2$, which is in the region slightly above the phase transition where we expect the negativity maximum. As shown in Fig.~\ref{fig:NegativityN2d4}, for $m$-independent dissipation the negativity decreases with $\gamma_C$ like for $N=5$ and $N=7$ (Fig.~\ref{fig:negativity}), but furthermore the negativity increases again for dominating $\gamma_C$ outside the parameter range shown in Fig.~\ref{fig:negativity}. Since the restriction $\rho_{\text{ss,sym}}$ to the symmetric subspace, $\rho_{\text{ss,sym}}=\mathcal{S}\rho_{\text{ss}}\mathcal{S}/\Tr[\mathcal{S}\rho_{\text{ss}}\mathcal{S}]$ with $\mathcal{S}=\sum_{J=3,1}\sum_{M=-J}^{J}\ket{J,M}\bra{J,M}$ 
here,
shows qualitatively the same behavior of negativity as the full steady state, the essence of this phenomenon should be understandable from investigating just $\rho_{\text{ss,sym}}$. As shown in the (blue) insets, the minimal negativity corresponds to strong couplings between $\ket{J=1,M=-1}$ and $\ket{J=3,M}$ and a comparably large weight on the $J=1$ subspace, whereas a smaller weight on $J=1$ (smallest $\gamma_C$ shown) as well as suppressed couplings between $\ket{J=1,M=-1}$ and $\ket{J=3,M}$ (largest $\gamma_C$, with even larger weight on $J=1$) are both linked to a larger negativity. Comparison with spin-ladder dissipation (shown in Fig.~\ref{fig:NegativityN2d4} in red) confirms this observation: There, the negativity increases monotonously with $\gamma_C$ and the largest negativity corresponds to suppressed couplings between $J=3$ and $J=1$ (as expected from the block-diagonal form of $J_-$). However, the relation between the matrix structure of $\rho_{\text{ss}}$ and the unusual behavior of negativity is still not fully clear and further research on other system sizes and also on more general dissipators would be needed for a deeper understanding.

\section{Proposal for Experimental Implementation}

\label{sec:Experiment}

\begin{figure}
	\includegraphics{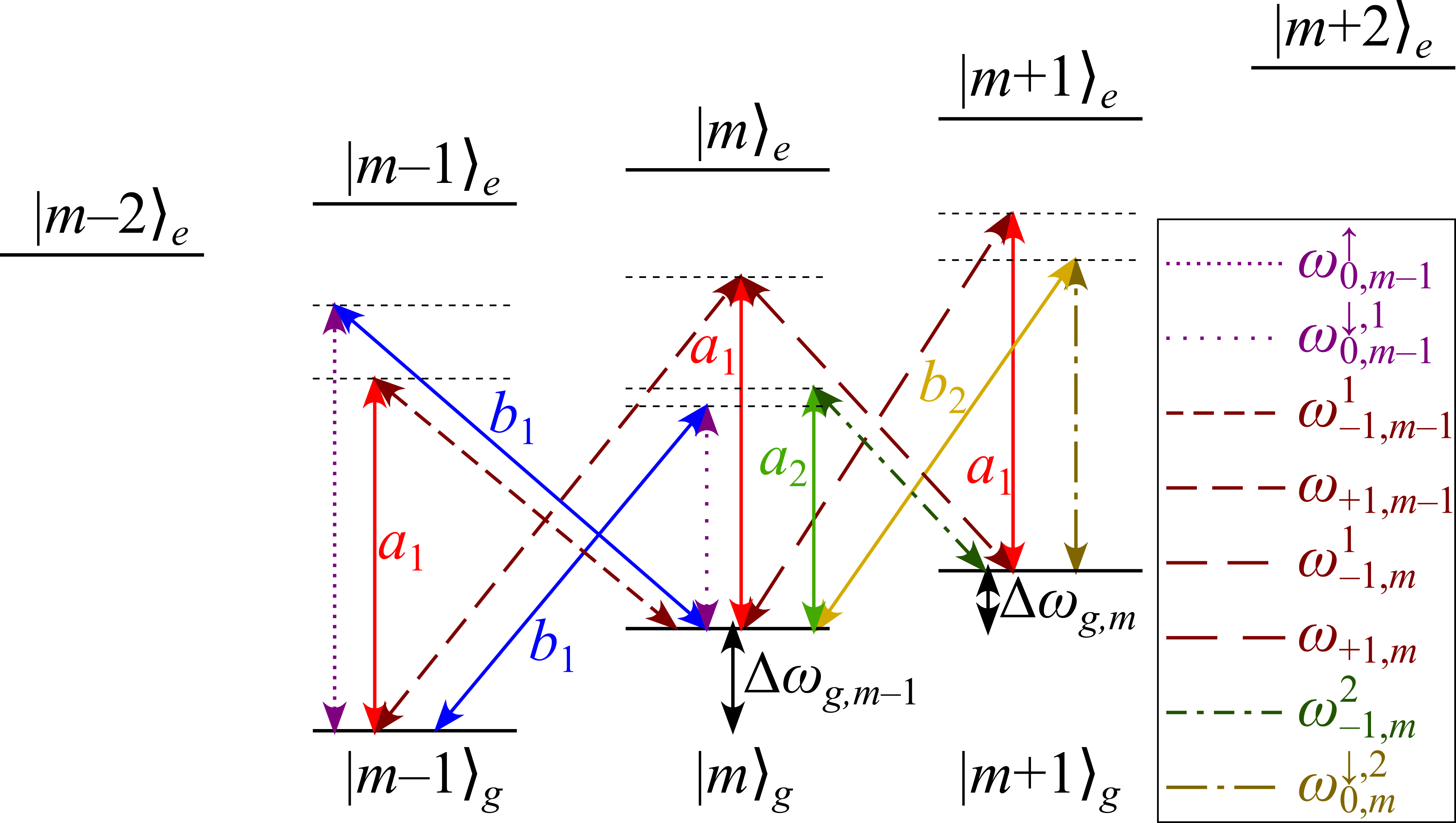}
	\caption{Example of level and excitation scheme for $d=3$ ground states and $d+2=5$ excited states. 
			The levels couple to four cavity modes $a_{1}$, $a_2$, $b_{1}$, $b_2$ (solid lines, with labels indicating the modes) and to a total of $6(d-1)=12$ driving fields (dotted, dash-dotted and dashed lines, with driving frequencies as indicated in the legend).
			For ease of visualization, the driving frequencies $\omega^\uparrow_{0,m}$, $\omega^\downarrow_{0,m}$, $\omega_{-1,m-1}^2$, $\omega_{0,m-1}^2$ are not shown, and the modes $a_2$, $b_1$, $b_2$ are shown only for a single pair of ground-state levels.
		Notations are explained in the main text. }
	\label{fig:SketchDoubleRaman}
\end{figure}

To complement our theoretical study, let us briefly discuss a possible experimental realization using cavity QED, based on ideas presented in Refs.~\cite{Dimer2007,Morrison2008,Morrison2008a,Huber2020}. 
The general scheme is sketched in Fig.~\ref{fig:SketchDoubleRaman}: $N$ identical atoms with a ground-state manifold of $d=2j+1$ levels $\ket{m}_g$ at energies $\omega_{g,m}$ ($m=-j,\ldots,j$) are coupled off-resonantly to $d+2$ excited-state levels $\ket{m'}_e$ at energies $\omega_{e,m'}$ ($m' = -j-1,\ldots,j+1$), where we set $\hbar=1$ for simplicity. The existence of the additional excited-state levels $m'=\pm(j+1)$ ensures that the two ground-state levels $m=\pm j$ can be coupled to the excited manifold in the same way as the other levels, and that the transition $\ket{m=0}_g\leftrightarrow \ket{m=0}_e$ is not forbidden, as it would be the case for identical $j$ in both manifolds \cite{CohenTannoudji,MetcalfBook}.
The frequencies $\Delta \omega_{g,m} = \omega_{g,m+1}-\omega_{g,m}$ are assumed to be distinct for different $m$, which could be achieved, e.g., 
by off-resonant driving fields coupling the ground-state manifold to another excited-state manifold \cite{MetcalfBook}. 

The transitions between ground and excited states are driven off-resonantly by $6(d-1)$ lasers of Rabi frequencies $\Omega_{0,\mu}^\uparrow$, $\Omega_{0,\mu}^{\downarrow,l}$, $\Omega_{+1,\mu}$, $\Omega_{-1,\mu}^l$
and corresponding driving frequencies $\omega_{0,\mu}^\uparrow$, $\omega_{0,\mu}^{\downarrow,l}$, $\omega_{+1,\mu}$, $\omega_{-1,\mu}^l$ ($l=1,2$, $\mu=-j,\ldots,j-1$), with polarizations such that the frequencies with first lower index $n$ couple only to the transitions $\ket{m}_g\leftrightarrow\ket{m+n}_e$.
Furthermore, the atomic transitions couple off-resonantly to
two pairs of cavity modes $a_{l}$, $b_{l}$ ($l=1,2$)
with frequencies $\omega_{a_{l}}$, $\omega_{b_l}$, polarized such that $a_{l}$ ($b_{l}$) are coupled only to the $\ket{m}_g\leftrightarrow\ket{m}_e$ ($\ket{m}_g\leftrightarrow\ket{m\pm 1}_e$) transitions, with coupling strengths $g_{0,m}^{l}$ ($g_{\pm 1,m}^{l}$). 

The frequencies of driving and cavity modes are assumed to fulfill the resonance conditions ($l=1,2$, $m=-j,\ldots,j-1$)
\begin{subequations}
	\label{eq:resonanceConditions}
	\begin{align}
		\omega_{a_l} &\approx \omega_{-1,m}^l+\Delta \omega_{g,m}, & \omega_{a_1} + \Delta \omega_{g,m} &\approx \omega_{+1,m},\\
		\omega_{b_l} &\approx \omega_{0,m}^{\downarrow,l} + \Delta \omega_{g,m}, & \omega_{b_1} + \Delta \omega_{g,m}&\approx \omega_{0,m}^\uparrow,
	\end{align}
	as sketched in Fig.~\ref{fig:SketchDoubleRaman}, and consequently also
	\begin{align}
		\omega_{+1,m} &\approx \omega_{-1,m'}^1+\Delta \omega_{g,m'} + \Delta \omega_{g,m},\\
		\omega_{0,m}^\uparrow &\approx \omega_{0,m'}^{\downarrow,1}+\Delta \omega_{g,m'} + \Delta \omega_{g,m},
	\end{align}
	for all $m,m'=-j,\ldots,j-1$.
\end{subequations}
Any combination of frequencies that is not constrained by Eq.~\eqref{eq:resonanceConditions} to be in resonance is assumed to be off-resonant. 

Adiabatic elimination \cite{Lugiato2015} of the excited-state manifold then yields the effective Hamiltonian
\begin{align}
	\nonumber
	H_\text{eff} ={}& \sum_{k=a_1,a_2,b_1,b_2} \left(\delta_k k^\dagger k - \sum_{m=-j}^j \zeta_{k,m} k^\dagger k S_{x;m,m}\right)\\
	\label{eq:effectiveHamiltonian}
	&+ \sum_{m=-j}^{j} \eps_m S_{x;m,m} - \sum_{k=a_1,a_2,b_1,b_2} \left(X_k k + X_k^{\dagger} k^\dagger\right) \\
	\nonumber
	&- \sum_{m=-j+1}^{j-1} \left(\xi_{x,m} S_{x;m+1,m-1} + \xi_{y,m} S_{y;m+1,m-1} \right).
\end{align}
Here, $X_k = N \sum_{m=-j}^{j-1} \left(\alpha_{k,m} S_{x;m+1,m} + \beta_{k,m} S_{y;m+1,m}\right)$ is a superposition of the operators $S_{x;m,n}$, $S_{y;m,n}$ introduced in Eq.~\eqref{eq:S-operators}, 
with coefficients
\begin{subequations}
\label{eq:alpha-beta-experiment}
\begin{align}
	\alpha_{a_l,m} ={}& \frac{\Omega_{-1,m}^{l*} g_{0,m}^l}{2 \Delta_m} + \frac{\Omega_{+1,m}^{l*} g_{0,m+1}^l}{2\Delta_{m+1}}, \\
	\alpha_{b_l,m} ={}& \frac{\Omega_{0,m}^{\downarrow,l*} g_{+1,m}^l}{2\Delta_{m+1}} + \frac{\Omega_{0,m}^{\uparrow,l*} g_{-1,m+1}^l}{2 \Delta_m}, \\
	\beta_{a_l,m} ={}& \i\left(\frac{\Omega_{-1,m}^{l*} g_{0,m}^l}{2 \Delta_m} - \frac{\Omega_{+1,m}^{l*} g_{0,m+1}^l}{2\Delta_{m+1}}\right), \\
	\beta_{b_l,m} ={}& \i\left( \frac{\Omega_{0,m}^{\downarrow,l*} g_{+1,m}^l}{2\Delta_{m+1}} - \frac{\Omega_{0,m}^{\uparrow,l*} g_{-1,m+1}^l}{2 \Delta_m}\right),
\end{align}
\end{subequations}
where $\Delta_m$ is of the order of the detuning between the transition frequencies and the driving frequencies and where $\Omega_{+1,m}^{2} = \Omega_{0,m}^{\uparrow,2} = 0$, consequently $\beta_{a_2,m} = \i \alpha_{a_2,m}$, $\beta_{b_2,m} = \i \alpha_{b_2,m}$.
Further details on the derivation of this equation can be found in Appendix~\ref{sec:details-experiment}, where, in Eq.~\eqref{eq:parameters-experiment}, also the other parameters of $H_\text{eff}$ are defined. 

The full system dynamics, including the decay of the cavity modes with dissipation rates $\kappa_k$ 
($k=a_1,a_2,b_1,b_2$) but neglecting individual dissipation for simplicity, is then described by the master equation
\begin{align}
	\nonumber
	\dot{\rho}_\text{full} ={}& -\i [H_\text{eff},\rho_\text{full}] \\
	&+ \sum_{k=a_1,a_2,b_1,b_2} \kappa_k \left(2 k \rho_\text{full} k^\dagger - \left\{k^\dagger k,\rho_\text{full}\right\}\right).
	\label{eq:atom-cavity-masterEq}
\end{align}
Assuming that the dynamics of the cavity modes alone is much faster than the coupled dynamics of atoms and modes according to Eq.~\eqref{eq:atom-cavity-masterEq} [bad cavity limit], we can adiabatically eliminate the cavity modes \cite{Azouit2017} (for details see Appendix~\ref{sec:details-experiment})
and arrive at the following master equation for just the atomic density matrix $\rho = \Tr_{\text{photons}}[\rho_{\text{full}}]$:
\begin{align}
	\dot{\rho} ={}& -\i [H,\rho] + \sum_{k=a_1,a_2,b_1,b_2} \frac{\kappa_k}{\kappa_k^2 + \delta_k^2}\left(2 X^{\dagger}_k \rho X_k - \left\{X_k X_k^\dagger, \rho\right\}\right)
	\label{eq:finalResultExperiment}
\end{align}
with 
\begin{align}
	\nonumber
	H ={}& \sum_{m=-j}^{j} \eps_m S_{x;m,m} - \sum_{k=a_1,a_2,b_1,b_2} 
	\frac{\delta_k}{\kappa_{k}^2+\delta_{k}^2} X_{k} X_{k}^\dagger\\
	&- \sum_{m=-j+1}^{j-1} \left(\xi_{x,m} S_{x;m+1,m-1} + \xi_{y,m} S_{y;m+1,m-1} \right).
\end{align}
Note that terms $\sim k^\dagger k S_{x;m,m}$ vanish in this limit, since the adiabatic elimination here effectively corresponds to an empty cavity.

By choosing Rabi frequencies, driving frequencies and  mode frequencies such that
\begin{enumerate}
	\item $\eps_m$ is independent of $m$ and can thus be discarded,
	\item $|\delta_{a_1}| \gg \kappa_{a_1}$, $|\delta_{b_1}| \gg \kappa_{b_1}$ and the corresponding contribution to dissipation is hence negligibly small, 
	\item $-\delta_{a_1}/(\delta_{a_1}^2+\kappa_{a_1}^2) = \delta_{b_1}/(\delta_{b_1}^2+\kappa_{b_1}^2) = \omega_c^{-2} V/(N j)$, $\alpha_{a_1,m} = \beta_{b_1,m} = A_{j,m} \omega_c$, $\beta_{a_1,m} = \alpha_{b_1,m} = 0$ with an arbitrary constant nonzero frequency $\omega_c$,
	\item $\alpha_{a_2,m} =  \ell_m^* \left(\gamma_C(\kappa_{a_2}^2 + \delta_{a_2}^2)/ N j \kappa_{a_2}\right)^{1/2}/2$ and $\alpha_{b_2,m} =  \ell_m^* \left(\gamma_C(\kappa_{b_2}^2 + \delta_{b_2}^2)/ N j \kappa_{b_2}\right)^{1/2}/2$, 
	\item $\delta_{a_2}/\kappa_{a_2} = -\delta_{b_2}/\kappa_{b_2}$, and
	\item $\xi_{x,m} = \xi_{y,m} = 0$,
\end{enumerate}
this effective model is identical to Eq.~\eqref{eq:master-eq}, except for the individual dissipation that we neglected here. Since $\alpha_{a_2,m}$, $\alpha_{b_2,m}$ can be tuned independently for each $m$, we can freely choose $\ell_m$ (within the ranges given by the accessible Rabi frequencies and detunings) and hence implement arbitrary collective dissipators $L_C$. Note that one might need to introduce further driving fields in order to jointly fulfill conditions~3 and~6.

As an example, we consider the transition between the $5S_{1/2}, F=2$ and $5P_{3/2}, F'=3$ hyperfine manifolds in \ce{^87Rb}, which has recently been employed as an experimentally accessible model system for the effects of multilevel systems on collective atom-cavity couplings \cite{Suarez2023}. 
We assume the driving frequencies to be larger than the transition frequencies by $\Delta\approx\SI{1}{\giga\hertz}$, 
a magnetic field of $\SI{5}{\milli\tesla}$ providing Zeeman splittings of \SI{35}{\mega\hertz}, $N\approx 10^4$ atoms like in Ref.~\cite{Suarez2023}, cavity decay rates $\kappa_k\approx\SI{10}{\kilo\hertz}$ (compare \cite{Klinner2006}, $k=a_1,a_2,b_1,b_2$), and coupling constants $g\approx\SI{210}{\kilo\hertz}$ (compare \cite{Suarez2023}). The coupling constants of each transition $\ket{m}_g\leftrightarrow\ket{n}_e$ are then given as $g c_{m,n}$, where $c_{m,n}$ is a Clebsch-Gordan coefficient \cite{MetcalfBook}.
By off-resonantly driving the transition between $5S_{1/2}, F=2$ and $5P_{1/2}, F'=1$, the level distances $\Delta \omega_{g,m}$ are modified by few \unit{\mega\hertz} and become distinct for different $m$. 
Furthermore, by tuning the offsets between driving frequencies and mode frequencies, we set $\delta_{a_l} = -\delta_{b_l} = \SI{102.9}{\kilo\hertz} \gg \kappa_{a_l},\kappa_{b_l}$ ($l=1,2$).
Choosing Rabi frequencies $\Omega_{-1,\mu=-2}^1 = \SI{6}{\mega\hertz}$, $\Omega_{-1,\mu=-2}^2 = \Omega_{0,\mu=-2}^{\downarrow,2}=\SI{50}{\mega\hertz}$ and appropriately adjusting the other Rabi frequencies in a range between $\approx\SI{4}{\mega\hertz}$ and $\approx\SI{50}{\mega\hertz}$ to fulfill conditions 3 and 4 (with $L_C = J_-$ or $L_C = \sum_{i=1}^N L^{(i)}_{\equiv}$), we obtain the interaction strength $|V|\approx\SI{12.7}{\kilo\hertz}$ and the collective dissipation rate $\gamma_C\approx\SI{25.8}{\kilo\hertz} \approx 2 |V|$. 
By tuning the Rabi frequencies from that configuration, one can then access the interaction- and dissipation-dominated regimes $\gamma_C < 2 |V|$ and $\gamma_C > 2 |V|$. 

\section{Conclusions}
\label{sec:conclusions}

We have thoroughly investigated 
a $d$-level generalization of the dissipative LMG model with individual and collective decay, in the mean-field limit of infinite particle numbers and numerically for finite system sizes. 
For any number of levels $d$, a dissipative phase transition has been identified as a function of interaction and dissipation, encompassing in particular the case of qubits studied earlier~\cite{Lee2013,Lee2014}.
The two phases are a symmetric, spin-$z$ polarized phase and a 
phase of broken symmetry, 
which hosts two static steady states in the presence of individual dissipation and an infinite number of oscillatory solutions in the case of only collective dissipation.

Under suitable parameter scaling, the position of the critical point can be made identical for all numbers $d$ of single-particle levels and independent of the exact form of the dissipators.
However, as $d$ increases, the critical exponents characterizing the behavior of spin expectation values at the phase transition reduce towards 0.
Furthermore, when $d\geq 4$ and the decay rates between adjacent single-particle levels $\ket{m}$ and $\ket{m+1}$ are identical for all $m$, the critical point transforms into a bistable region, in which both phases are present. 

The numerically computed spectrum of the Liouvillian confirms the existence of the phase transition and the spontaneous symmetry breaking accompanying it: At the transition to the 
broken-symmetry 
phase, the associated gap in the Liouvillian spectrum closes and remains closed within that phase. In addition, the bistable region is signalled by a vanishing gap in the Liouvillian spectrum of the symmetric subspace (with respect to the model's $\Z_2$ symmetries).

The purity of the steady state reveals that the two phases can be characterized as almost pure (spin-$z$ polarized phase) and highly mixed 
(broken-symmetry 
phase), where the bistable region is found to belong to the highly mixed phase. As with other open systems \cite{Schneider2002,GonzalezTudela2013,Lee2013a,Lee2014a,Wolfe2014,Barberena2019,Wang2023e}, the dissipative phase transition is accompanied by a maximum of steady-state entanglement.
Surprisingly, the interplay of individual and collective terms can lead to stronger entanglement than collective terms alone, an effect that is only present for $d\geq 4$. 

We have thus found that the $d$-level generalization of the dissipative LMG model not only allows for a richer phase diagram, with a bistable region not present in the two-level case, but also for new effects of dissipation onto steady-state entanglement.
In particular this last point would deserve further investigation in future research. Understanding in detail the reasons for the suppression and enhancement of entanglement by the individual or collective nature of dissipation could lead to new schemes of preparing entangled steady states.
Furthermore, as spin squeezing is related to entanglement \cite{MaSpinSqueezing}, it might show a similar dependence on individual and collective dissipation as the entanglement negativity investigated in this work. Consequently, individual dissipation could, for particular dissipators, ease the preparation of spin-squeezed steady states, a finding that could widely be applied in the context of quantum metrology.


\begin{acknowledgments}
	We thank J.~Louvet and B.~Debecker for fruitful discussions. This project (EOS 40007526) has received funding from the FWO and F.R.S.-FNRS under the Excellence of Science (EOS) programme.
\end{acknowledgments}


\appendix

\section{Mean-field Equations of Motion}
\label{sec:details-mean-field}

When the time evolution of the density matrix $\rho$ is given by Eq.~\eqref{eq:master-eq}, the expectation value $\left<O\right> = \operatorname{Tr}[O\rho]$ of a time-independent Hermitian operator $O$ evolves according to 
\begin{equation}
	\begin{aligned}
         \frac{d\left<O\right>}{dt}
		={}& \i \left<[H,O]\right> + \frac{\gamma_I}{2 j}\sum_{i=1}^N \left< \left[ \left(L^{(i)}\right)^\dagger,O \right] L^{(i)} + \text{H.c.} 
		\right> \\
		&+ \frac{\gamma_C}{2Nj} \left< [L_C^\dagger,O] L_C + \text{H.c.} 
		\right>,
	\end{aligned}
\end{equation}
where $\text{H.c.}$ stands for the Hermitian conjugate of the preceding term, i.e., for $\left(\left[\left(L^{(i)}\right)^\dagger,O \right] L^{(i)}\right)^\dagger = \left(L^{(i)}\right)^\dagger \left[O,L^{(i)}\right]$ and $\left([L_C^\dagger,O] L_C\right)^\dagger = L_C^\dagger [O,L_C]$ here.
Applying the commutation relations 
\begin{subequations}
	\label{eq:commutators}
	\begin{align}
		&\begin{aligned}
			-2\i N[S_{x;m,n},S_{x;o,p}] ={}& \delta_{n,o} S_{y;m,p} + \delta_{n,p} S_{y;m,o} 
			\\
			&+ \delta_{m,o} S_{y;n,p} + \delta_{m,p} S_{y;n,o},
		\end{aligned}\\
		&\begin{aligned}
			-2\i N[S_{y;m,n},S_{y;o,p}] ={}& - \delta_{n,o} S_{y;m,p} + \delta_{n,p} S_{y;m,o} \\
			&+ \delta_{m,o} S_{y;n,p} - \delta_{m,p} S_{y;n,o},\\
		\end{aligned}\\
		&\begin{aligned}
			-2\i N[S_{x;m,n},S_{y;o,p}] ={}& - \delta_{n,o} S_{x;m,p} + \delta_{n,p} S_{x;m,o} \\
			&- \delta_{m,o} S_{x;n,p} + \delta_{m,p} S_{x;n,o}
		\end{aligned}
	\end{align}
\end{subequations}
to the unitary part yields
\begin{subequations}
	\label{eq:mean-field-unitary}
	\begin{align}
		\i[H,S_{x;m,n}] ={}& \frac{V}{2j}  \left(\left\{\frac{J_x}{N}, \AA^{y;mn}_{+1,+1}\right\} + \left\{\frac{J_y}{N},\AA^{x;mn}_{-1,-1}\right\}\right),\\
		\i[H,S_{y;m,n}] ={}& \frac{V}{2j} \left(\left\{\frac{J_y}{N},\AA^{y;mn}_{-1,+1}\right\} - \left\{\frac{J_x}{N}, \AA^{x;mn}_{+1,-1}\right\}\right),
	\end{align}
\end{subequations}
where we defined the short-hand notation
\begin{align}
	\begin{aligned}
		\AA_{\sigma_1,\sigma_2}^{\alpha;mn} ={}& A_{j,n-1} S_{\alpha;m,n-1} + \sigma_1\sigma_2 A_{j,m-1} S_{\alpha;n,m-1}\\
		&+\sigma_1 A_{j,n} S_{\alpha;m,n+1} + \sigma_2 A_{j,m} S_{\alpha;n,m+1}
	\end{aligned}
\end{align} 
and used that $J_x/N = \sum_{m=-j}^{j-1} A_{j,m} S_{x;m+1,m}$ and $J_y/N = \sum_{m=-j}^{j-1} A_{j,m} S_{y;m+1,m}$. Similarly, expressing $L_C$ as $L_C = N\sum_{m=-j}^{j-1} \ell_m \left(S_{x;m+1,m} - \i S_{y;m+1,m}\right)$ and applying the commutation relations gives
\begin{widetext}
	\begin{subequations}
		\label{eq:mean-field-collective}
		\begin{align}
			&\begin{aligned}
				[L_C^\dagger,S_{x;m,n}]L_C + \text{H.c.} ={}&
				\frac{N}{2}\sum_{o=-j}^{j-1} \left(-\left\{\RR^{y;mno}_{+1,+1}+\II^{x;mno}_{-1,-1},S_{y;o+1,o}\right\}
				- \left\{\RR^{x;mno}_{-1,-1}-\II^{y;mno}_{+1,+1},S_{x;o+1,o}\right\} \right.\\
				&+ \left. \i \left[\RR^{x;mno}_{-1,-1}-\II^{y;mno}_{+1,+1},S_{y;o+1,o}\right] - \i \left[\RR^{y;mno}_{+1,+1}+\II^{x;mno}_{-1,-1},S_{x;o+1,o}\right]\right),
			\end{aligned}\\
			&\begin{aligned}
				[L_C^\dagger,S_{y;m,n}]L_C + \text{H.c.} ={}& \frac{N}{2}\sum_{o=-j}^{j-1} \left(-\left\{\RR^{y;mno}_{-1,+1}+\II^{x;mno}_{+1,-1},S_{x;o+1,o}\right\}
				+ \left\{\RR^{x;mno}_{+1,-1} - \II^{y;mno}_{-1,+1}, S_{y;o+1,o} \right\} \right.\\
				&+ \left.\i \left[\RR^{x;mno}_{+1,-1} - \II^{y;mno}_{-1,+1},S_{x,o+1,o}\right]
				+\i \left[\RR^{y;mno}_{-1,+1} + \II^{x;mno}_{+1,-1},S_{y;o+1,o}\right]\right).
			\end{aligned}
		\end{align}
	\end{subequations}
\end{widetext}
Here, 
we introduced the short-hand notations
\begin{subequations}
	\begin{align}
		&\begin{aligned}
			\RR^{\alpha;mno}_{\sigma_1,\sigma_2} ={}& \Re[\ell_{n-1}^*\ell_{o}] S_{\alpha;m,n-1} + \sigma_1\sigma_2 \Re[\ell_{m-1}^*\ell_o] S_{\alpha;n,m-1}\\
			&+\sigma_1 \Re[\ell_n^*\ell_o] S_{\alpha;m,n+1}  + \sigma_2 \Re[\ell_{m}^*\ell_o] S_{\alpha;n,m+1},
		\end{aligned}\\
		&\begin{aligned}
			\II^{\alpha;mno}_{\sigma_1,\sigma_2} ={}& \Im[\ell_{n-1}^*\ell_{o}] S_{\alpha;m,n-1} + \sigma_1\sigma_2 \Im[\ell_{m-1}^*\ell_o] S_{\alpha;n,m-1}\\
			&+ \sigma_1 \Im[\ell_n^*\ell_o] S_{\alpha;m,n+1} + \sigma_2 \Im[\ell_{m}^*\ell_o] S_{\alpha;n,m+1}.
		\end{aligned}
	\end{align}
\end{subequations}
For the individual dissipators, note that 
\begin{subequations}
	\begin{align}
		&\begin{aligned}
			[\ket{o+1}_i\bra{o}_i,S_{x;m,n}] ={}& \frac{\delta_{o,m}}{2N} \ket{o+1}_i\bra{n}_i + \frac{\delta_{o,n}}{2N} \ket{o+1}_i\bra{m}_i \\
			&- \frac{\delta_{o+1,n}}{2N} \ket{m}_i\bra{o}_i - \frac{\delta_{o+1,m}}{2N} \ket{n}_i\bra{o}_i,\\
		\end{aligned}\\
		&\begin{aligned}
			[\ket{o+1}_i\bra{o}_i,S_{y;m,n}] ={}& \frac{\delta_{o,m}}{2 \i N}\ket{o+1}_i\bra{n}_i - \frac{\delta_{o,n}}{2 \i N}\ket{o+1}_i\bra{m}_i \\
			&- \frac{\delta_{o+1,n}}{2 \i N}\ket{m}_i\bra{o}_i + \frac{\delta_{o+1,m}}{2 \i N}\ket{n}_i\bra{o}_i.
		\end{aligned}
	\end{align}
\end{subequations}
Inserting these expressions into $\left[\left(L^{(i)}\right)^\dagger,S_{\alpha;m,n} \right]L^{(i)}$ with $\alpha = x,y$ and summing over $i$ yields
\begin{subequations}
	\label{eq:individual_diss}
	\begin{align}
		&\begin{aligned}
			\sum_{i=1}^N \left[\left(L^{(i)}\right)^\dagger,S_{x;m,n} \right] L^{(i)} + \text{H.c.} ={}&
			2 \bigg(\!\Re\left[\ell_{m}^*\ell_n \right]  S_{x;m+1,n+1}\\
			&- \Im\left[ \ell_{m}^*\ell_n \right]  S_{y;m+1,n+1}\\
			&- \frac{|\ell_{n-1}|^2 + |\ell_{m-1}|^2}{2} S_{x;m,n}\bigg),\\
		\end{aligned}\\
		&\begin{aligned}
			\sum_{i=1}^N \left[\left(L^{(i)}\right)^\dagger,S_{y;m,n} \right] L^{(i)} + \text{H.c.} ={}& 
			2 \bigg(\!\Re\left[\ell_{m}^*\ell_n \right] S_{y;m+1,n+1}\\
			&+ \Im\left[ \ell_{m}^*\ell_n \right] S_{x;m+1,n+1}\\
			&- \frac{|\ell_{n-1}|^2 + |\ell_{m-1}|^2}{2} S_{y;m,n}\bigg).
		\end{aligned}
	\end{align}
\end{subequations}

Since the commutator $[S_{\alpha;m,n},S_{\beta;o,p}]$ vanishes as $N\to\infty$ [compare Eq.~\eqref{eq:commutators}], we can treat the $S$ operators as real numbers in the limit $N\to\infty$. In particular we approximate $\left<\{S_{\alpha;m,n},S_{\beta;o,p}\}\right> \approx 2 \left<S_{\alpha;m,n}\right>\left<S_{\beta;o,p}\right>$ and consequently also
\begin{subequations}
	\begin{align}
		\left<\left\{\frac{J_\alpha}{N},\AA^{\beta;mn}_{\sigma_1,\sigma_2}\right\}\right> &\approx 2\frac{\left<J_\alpha\right>}{N} \left<\AA^{\beta;mn}_{\sigma_1,\sigma_2}\right>,\\
		\left<\left\{\RR^{\alpha;mno}_{\sigma_1,\sigma_2},S_{\beta,o+1,o}\right\}\right> &\approx 2 \left<\RR^{\alpha;mno}_{\sigma_1,\sigma_2}\right> \left<S_{\beta,o+1,o}\right>, \\
		\left<\left\{\II^{\alpha;mno}_{\sigma_1,\sigma_2},S_{\beta,o+1,o}\right\}\right> &\approx 2 \left<\II^{\alpha;mno}_{\sigma_1,\sigma_2}\right> \left<S_{\beta,o+1,o}\right>,
	\end{align}
\end{subequations}
while we set all commutators to 0. This yields a closed set of differential equations for $\left<S_{\alpha;m,n}\right>$, with two quadratic terms stemming from the unitary time evolution and from the collective dissipation [$f_{\alpha;m,n}$ and $h_{\alpha;m,n}$ in Eq.~\eqref{eq:mean-field}], and a linear term due to the individual dissipation [$g_{\alpha;m,n}$ in Eq.~\eqref{eq:mean-field}].

\section{Fixed Points and Their Stability}
\label{sec:details-fixed-points}

Simple fixed points of the mean-field equations derived in the preceding appendix can be found by individually setting each contribution (unitary part, individual and collective dissipation) to 0. 
When $\gamma_I \neq 0$, the individual dissipation vanishes only if $\left<S_{x;m,n}\right> = \left<S_{y;m,n}\right> = 0$ for all $m,n=-j,\ldots,j$ with $\ell_{m-1} \neq 0$ or $\ell_{n-1} \neq 0$.
This can be seen by an induction over $n$, starting from $n$ with $\ell_n= 0$ (e.g., $n=j$). 
For the most general form of $L^{(i)}$, thus only $\left<S_{x;-j,-j}\right>$ and $\left<S_{y;-j,-j}\right>$ are allowed to be nonzero.
Noting that $\left<S_{y;-j,-j}\right> = 0$ by definition and that the normalization condition $\sum_{m=-j}^j \left<S_{x;m,m}\right> = 1$ fixes $\left<S_{x;-j,-j}\right> = 1$, one obtains exactly the spin-$z$ polarized steady state defined in Eq.~\eqref{eq:southpole}.
It is easy to see that this choice of $\left<S_{\alpha;m,n}\right>$ also makes the unitary part and the collective dissipation vanish and thus constitutes a fixed point for all values of $d, V, \gamma_{I}, \gamma_C$ and $L^{(i)}$.

To study the stability of a fixed point $s$, one computes the eigenvalues of the Jacobian matrix $\mathcal{J}(s)$ \cite{Ott2002,Strogatz2015}, i.e., the 
matrix of partial derivatives $\mathcal{J}_{\beta;o,p}^{\alpha;m,n}(s) = \frac{\del F_{\alpha;m,n}}{\del \left<S_{\beta;o,p}\right>}(s)$ at $s$, where $F_{\alpha;m,n}\left(\left<S_{\beta;o,p}\right>\right) = V \, f_{\alpha;m,n}\!\left(\left<S_{\beta;o,p}\right>\right) + \gamma_I \, g_{\alpha;m,n}\!\left( \left<S_{\beta;o,p}\right>\right) + \gamma_C \,  h_{\alpha;m,n}\!\left(\left<S_{\beta;o,p}\right>\right)$ is the right-hand side of Eq.~\eqref{eq:mean-field}. 
After eliminating $\left<S_{x;-j,-j}\right>$ via the condition $\sum_{m=-j}^j \left<S_{x;m,m}\right> = 1$, 
one obtains for the spin-$z$ polarized steady state
\begin{widetext}
	\begin{subequations}
		\begin{align}
			\mathcal{J}_{x;o,p}^{x;m,n}(s) ={}& -\frac{\gamma_{C}}{2 j}\Re[\ell_{-j}^*\ell_p]\delta_{o,p+1}\delta_{-j+1,m}\delta_{-j,n} 
			+ \frac{\gamma_I}{j}\left(\Re[\ell_m^*\ell_n]\delta_{m+1,o}\delta_{n+1,p} - \frac{|\ell_{m-1}|^2 + |\ell_{n-1}|^2}{2} \delta_{m,o}\delta_{n,p}\right),\\
			\mathcal{J}_{y;o,p}^{x;m,n}(s) ={}& \left(\frac{V A_{j,p} A_{j,-j}}{j} - \frac{\gamma_C}{2 j}\Im[\ell_{-j}^* \ell_p] \right)\delta_{o,p+1}\delta_{-j+1,m}\delta_{-j,n} 
			- \frac{\gamma_I}{j} \Im[\ell_m^*\ell_n] \delta_{m+1,o}\delta_{n+1,p},\\
			\mathcal{J}_{x;o,p}^{y;m,n}(s) ={}& \left(\frac{V A_{j,p} A_{j,-j}}{j} + \frac{\gamma_C}{2 j}\Im[\ell_{-j}^* \ell_p] \right)\delta_{o,p+1}\delta_{-j+1,m}\delta_{-j,n} 
			+ \frac{\gamma_I}{j} \Im[\ell_m^*\ell_n] \delta_{m+1,o}\delta_{n+1,p},\\
			\mathcal{J}_{y;o,p}^{y;m,n}(s) ={}& - \frac{\gamma_{C}}{2 j}\Re[\ell_{-j}^*\ell_p]\delta_{o,p+1}\delta_{-j+1,m}\delta_{-j,n} 
			+ \frac{\gamma_I}{j}\left(\Re[\ell_m^*\ell_n]\delta_{m+1,o}\delta_{n+1,p} - \frac{|\ell_{m-1}|^2 + |\ell_{n-1}|^2}{2} \delta_{m,o}\delta_{n,p}\right).
		\end{align}
	\end{subequations}
\end{widetext}
When the rows and columns of $\mathcal{J}(s)$ 
are sorted into groups of identical $m-n$  or $o-p$, respectively, within which the respective index $m$ or $o$ increases monotonically, 
the Jacobian matrix is triangular except for a block of size $2\times 2$, which reads
\begin{align}
	\begin{pmatrix}
		-\frac{|\ell_{-j}|^2(\gamma_I + \gamma_C)}{2j} & 2 V\\
		2 V & -\frac{|\ell_{-j}|^2(\gamma_I + \gamma_C)}{2j}
	\end{pmatrix},
\end{align}
and thus has eigenvalues $-(|\ell_{-j}|^2(\gamma_I + \gamma_C))/(2j) \pm 2V$.
The other eigenvalues of $\mathcal{J}(s)$ are the remaining diagonal elements: the $-1 + d(d-1)/2$ doubly degenerate values $- \gamma_I \left( |\ell_{m-1}|^2 + |\ell_{n-1}|^2 \right)/2 j$, with $m = -j + 2,\ldots, j$, $n = -j,\ldots,m-1$, $\ell_{-j-1}=0$, and the $d-1$ non-degenerate values $- \gamma_I |\ell_{m-1}|^2/j$, with $m=-j+1,\ldots,j$.
For $|\ell_{-j}|^2(\gamma_I + \gamma_C) > 4 j |V|$, all eigenvalues are negative and the spin-$z$-polarized steady state is thus a stable fixed point, whereas for $|\ell_{-j}|^2(\gamma_I + \gamma_C) < 4j|V|$  one of the eigenvalues in the $2\times 2$ block becomes positive and consequently the fixed point becomes unstable.

From Eqs.~\eqref{eq:mean-field-unitary} and~\eqref{eq:mean-field-collective}, it is clear that the unitary part and the collective dissipation vanish also at any other point that fulfills $\left<S_{x;m+1,m}\right> = \left<S_{y;m+1,m}\right> = 0$ for all $m=-j,\ldots,j-1$. Consequently, each such point is a fixed point for $\gamma_I = 0$. This includes, for instance, the diagonal states, where only the terms $\left<S_{x;m,m}\right>$, $m=-j,\ldots,j$, are allowed to be nonzero. The Jacobian at such a diagonal state reads, with $\mathcal{S}_{n} = \left<S_{x;n+1,n+1}\right> - \left<S_{x;n,n}\right>$,
\begin{subequations}
	\begin{align}
		\mathcal{J}_{x;o,p}^{x;m,n}(s) ={}& \frac{\gamma_{C}}{2 j}\Re[\ell_{n}^*\ell_p]
		\mathcal{S}_{n}
		\delta_{o,p+1}\delta_{m,n+1},\\
		\mathcal{J}_{y;o,p}^{x;m,n}(s) ={}& \left(-\frac{V A_{j,p} A_{j,n}}{j} + \frac{\gamma_C}{2 j}\Im[\ell_{n}^* \ell_p] \right)
		\mathcal{S}_{n}
		\delta_{o,p+1}\delta_{m,n+1},\\
		\mathcal{J}_{x;o,p}^{y;m,n}(s) ={}& -\left(\frac{V A_{j,p} A_{j,n}}{j} + \frac{\gamma_C}{2 j}\Im[\ell_{n}^* \ell_p] \right)
		\mathcal{S}_{n}
		\delta_{o,p+1}\delta_{m,n+1},\\
		\mathcal{J}_{y;o,p}^{y;m,n}(s) ={}& \frac{\gamma_{C}}{2 j}\Re[\ell_{n}^*\ell_p]
		\mathcal{S}_{n}
		\delta_{o,p+1}\delta_{m,n+1},
	\end{align}
\end{subequations}
i.e., it vanishes except for a block of dimension \mbox{$2(d-1)\times 2(d-1)$}. 
To get a basic understanding of the stability criteria for these states, we study a simple diagonal state with $\left<S_{x;m+1,m+1}\right> = \left<S_{x;m,m}\right>$ for $m=-j,\ldots,j-1$ except for one specific $m=m_0$. The Jacobian has then only two nonzero eigenvalues, which read
\begin{align}
	\frac{\mathcal{S}_{m_0}}{2j} \left(\left|\ell_{m_0}\right|^2 \gamma_C \pm 2 A_{j,m_0}^2 |V| \right).
\end{align}
This diagonal state is thus stable if and only if $\mathcal{S}_{m_0} < 0$ (which is equivalent to $\left<S_{x;m_0+1,m_0+1}\right> < \left<S_{x;m_0,m_0}\right>$) and $\left|\ell_{m_0}\right|^2 \gamma_C > 2 A_{j,m_0}^2 |V|$, i.e., a similar stability criterion as for the spin-$z$ polarized state, but with dependence on $\ell_{m_0}$ instead of $\ell_{-j}$. For more general diagonal states, where the Jacobian is a function of $\ell_m$ and $\mathcal{S}_m$ for all $m$, we expect also the stability to depend on all the $\ell_m$ and $\mathcal{S}_m$.

As discussed in Sec.~\ref{sec:mean-field}, further fixed points may in general emerge, for which the unitary part and the collective and individual dissipation do not vanish separately and the fixed points depend explicitly on $L^{(i)}$, $V$, $\gamma_C$ and $\gamma_{I}$. 
For $d=2$, these fixed points can be found analytically and are given in Sec.~\ref{sec:Qubits}. The corresponding eigenvalues of the Jacobian matrix for $\gamma_I\neq 0$ are $-(4|V|\gamma_I)/(2|V|-\gamma_C)$ and $-\gamma_I \pm \sqrt{\gamma_I\left(5\gamma_I + 4\gamma_C - 8|V|\right)}$, whose real part is negative as long as $\gamma_I + \gamma_C < 2|V|$ and turns positive for (at least) one eigenvalue when $\gamma_I + \gamma_C > 2|V|$. Hence, these fixed points are stable exactly for $\gamma_I + \gamma_C < 2|V|$, which is also the range of parameters for which the corresponding values of $\left<J_\alpha\right>$, $\alpha=x,y$, are real and thus physically meaningful.
For $\gamma_I = 0$, the eigenvalues of the Jacobian matrix are $\pm \sqrt{2 \gamma_{C}^2 - 8 V^2}$ (note that there are only two eigenvalues due to the condition $X^2+Y^2+Z^2 = 1$), which are imaginary for $\gamma_C < 2|V|$ and one of which becomes positive for $\gamma_C > 2|V|$. The corresponding fixed points are thus centers for $\gamma_C < 2 |V|$, which is, again, also the range of physically meaningful values of $\left<J_\alpha\right>$, $\alpha=x,y$, and they are unstable for $\gamma_C > 2 |V|$.
For $d>2$, these fixed points and their stability are calculated numerically, as described in Sec.~\ref{sec:Qudits}.

\section{Details of the Experimental Proposal}
\label{sec:details-experiment}

As sketched in Fig.~\ref{fig:SketchDoubleRaman}, we consider $N$ identical atoms labelled by $i=1,\ldots ,N$, with $d$ ground states at energies $\omega_{g,m}$ ($m=-j,\ldots,j$) and $d+2$ excited states at energies $\omega_{e,m}$ ($m = -j-1,\ldots,j+1$), coupled via $6(d-1)$ driving fields and $4$ cavity modes. 
We first move into the interaction picture with respect to the Hamiltonian
\begin{align}
	\begin{aligned}
		H_0 ={}& \sum_{i=1}^{N}\left[\sum_{m=-j}^{j}\omega_{g,m}'\ket{m}_{g,i}\bra{m}_{g,i} + \sum_{m=-j-1}^{j+1} \omega_{e,m}'\ket{m}_{e,i}\bra{m}_{e,i}\right]\\
		&+\sum_{k=a_1,a_2,b_1,b_2}\omega_k' k^\dagger k,
	\end{aligned}
\end{align}
where $\omega_{g,m}'$ and $\omega_k'$ are energies close to $\omega_{g,m}$ and $\omega_k$, respectively, chosen such that 
\begin{align}
	\label{eq:primedCavitiesA}
	\begin{aligned}
		\omega_{a_l}' - \omega_{-1,m}^l &= \omega_{b_l}' - \omega_{0,m}^{\downarrow,l}\\
		&= \omega_{g,m+1}' - \omega_{g,m}'
	\end{aligned}
	 \qquad (l=1,2)
\end{align}
and
\begin{align}
	\label{eq:resonanceDefinition}
	\begin{aligned}
		\omega_{0,m}^\uparrow - \omega_{0,m'}^{\downarrow,1} &= \omega_{+1,m} - \omega_{-1,m'}^1 \\
		&= \omega_{g,m+1}' + \omega_{g,m'+1}' - \omega_{g,m}' - \omega_{g,m'}'
	\end{aligned}
\end{align}
for all $m, m' = -j,\ldots,j-1$.
The energy difference $\omega_{e,m}'-\omega_{g,m}$ (for $m=\pm (j+1)$, the difference $\omega_{e,\pm(j+1)}'-\omega_{g,\pm j}$) is of the order of the driving frequencies and we assume that $\Delta_m = \omega_{e,m} - \omega_{e,m}'$ is much larger than the Rabi frequencies, the coupling strengths and the energy differences $\omega_{g,m}-\omega_{g,m}'$ and $\omega_k-\omega_k'$ ($m=-j,\ldots,j$, $k=a_{1},a_2,b_{1},b_2$). With a rotating-wave approximation, the Hamiltonian in the interaction picture is then 
\begin{subequations}
	\begin{align}
		H &= H_\text{a} + H_\text{ph} + H_\text{a-ph} + H_\text{d},
	\end{align}
	with
	\allowdisplaybreaks
	\begin{widetext}
		\begin{align}
			H_\text{a} ={}& \sum_{i=1}^N\left(\sum_{m=-j}^{j}\left(\omega_{g,m} - \omega_{g,m}'\right) \ket{m}_{g,i}\bra{m}_{g,i} + \sum_{m=-j-1}^{j+1} \Delta_m \ket{m}_{e,i}\bra{m}_{e,i}\right),\\
			H_\text{ph} ={}& \sum_{k=a_1,a_2,b_1,b_2}\left(\omega_k - \omega_k'\right) k^\dagger k,\\
			\nonumber
			H_\text{a-ph} ={}& \sum_{i=1}^{N}\sum_{m=-j}^{j}\sum_{l=1,2}\left[e^{\i(\omega_{e,m}' - \omega_{g,m}' - \omega_{a_l}') t} g_{0,m}^l a_l \ket{m}_{e,i}\bra{m}_{g,i} \right.
			+ e^{\i(\omega_{e,m+1}' - \omega_{g,m}' - \omega_{b_l}') t} g_{+1,m}^l b_l \ket{m+1}_{e,i}\bra{m}_{g,i}\\
			&+\left. e^{\i(\omega_{e,m-1}' - \omega_{g,m}' - \omega_{b_l}') t} g_{-1,m}^l b_l \ket{m-1}_{e,i}\bra{m}_{g,i} \right] + \text{H.c.},\\
			\nonumber
			H_\text{d} ={}& \sum_{i=1}^{N}\sum_{m=-j}^{j}\sum_{m'=-j}^{j-1}\left[\frac{e^{\i(\omega_{e,m}'-\omega_{g,m}')t}\left(\Omega_{0,m'}^\uparrow e^{-\i\omega_{0,m'}^\uparrow t} + \Omega_{0,m'}^{\downarrow,1} e^{-\i\omega_{0,m'}^{\downarrow,1} t} + \Omega_{0,m'}^{\downarrow,2} e^{-\i\omega_{0,m'}^{\downarrow,2} t}\right)}{2}
			\ket{m}_{e,i}\bra{m}_{g,i} 
			\right.\\
			&+ \frac{\Omega_{+1,m'} e^{\i(\omega_{e,m+1}' - \omega_{g,m}' - \omega_{+1,m'}) t}}{2} \ket{m+1}_{e,i}\bra{m}_{g,i}\\
			\nonumber
			&+\left. \frac{e^{\i(\omega_{e,m-1}'-\omega_{g,m}')t}\left(\Omega_{-1,m'}^{1} e^{-\i\omega_{-1,m'}^{1} t} + \Omega_{-1,m'}^{2} e^{-\i\omega_{-1,m'}^{2} t}\right)}{2}\ket{m-1}_{e,i}\bra{m}_{g,i} \right] + \text{H.c.},
		\end{align}
	\end{widetext}
\end{subequations}
where H.c.\ is the Hermitian conjugate.

Since $\Delta_m$ is much larger than the other frequencies, the excited states are only populated virtually. We can thus restrict to the subspace of maximally one excitation and adiabatically eliminate the excited-state manifold \cite{Lugiato2015}. With $P$ and $Q$  being the projectors onto the subspaces of zero and one excitation, respectively, adiabatic elimination corresponds to 
the effective ground-state Hamiltonian
\begin{align}
	H_\text{eff} = PHP - PHQ(QHQ)^{-1}QHP,
\end{align}
with $(QHQ)^{-1}\approx \sum_{i=1}^{N}\sum_{m=-j-1}^{j+1}\Delta_m^{-1}\ket{m}_{e,i}\bra{m}_{e,i}$ (neglecting small contributions to $QHQ$ from the cavity modes and the ground states). 
Inserting our specific model and neglecting any off-resonant contributions according to Eq.~\eqref{eq:resonanceConditions} yields Eq.~\eqref{eq:effectiveHamiltonian} with $\alpha_k$, $\beta_k$ given in Eq.~\eqref{eq:alpha-beta-experiment} and further parameters
\begin{subequations}
	\label{eq:parameters-experiment}
	\begin{align}
		\delta_k ={}& \omega_{k} - \omega_k' - N \delta_{k}^+, \\
		\delta_{a_{l}}^+ ={}& \frac{1}{d}\sum_{m=-j}^{j} \frac{\left|g_{0,m}^{l}\right|^2}{\Delta_{m}},\\
		\delta_{b_l}^+ ={}& \frac{1}{d} \sum_{m=-j}^{j} \left(\frac{\left|g_{+1,m}^{l}\right|^2}{\Delta_{m+1}} + \frac{\left|g_{-1,m}^{l}\right|^2 }{\Delta_{m-1}}\right),\\
		\frac{\zeta_{a_l,m}}{N} ={}& \frac{\left|g_{0,m}^l\right|^2}{\Delta_{m}} - \delta_{a_l}^+, \\
		\frac{\zeta_{b_l,m}}{N} ={}& \frac{\left|g_{+1,m}^l\right|^2}{\Delta_{m+1}} + \frac{\left|g_{-1,m}^l\right|^2 }{\Delta_{m-1}} - \delta^{+}_{b_l},
	\end{align}
	\begin{align}
		\nonumber
		\frac{\eps_m}{N} ={}& \left(\omega_{g,m} - \omega_{g,m}' \right) - \sum_{m' = -j}^{j-1} \left(\frac{\left|\Omega_{0,m'}^\uparrow\right|^2 + \left|\Omega_{0,m'}^{\downarrow,1}\right|^2 + \left|\Omega_{0,m'}^{\downarrow,2}\right|^2}{4\Delta_m} \right.\\
		&+ \left.\frac{\left|\Omega_{+1,m'}\right|^2}{4\Delta_{m+1}} + \frac{\left|\Omega_{-1,m'}^1\right|^2 + \left|\Omega_{-1,m'}^2\right|^2}{4\Delta_{m-1}} \right),
	\end{align}
	\begin{align}
		\xi_{x,m} ={}& \xi_m + \xi_m^*,
		\qquad \xi_{y,m} ={} \i \left( \xi_m - \xi_m^* \right),\\
		\frac{\xi_m}{N} ={}& \frac{\Omega_{-1,m-1}^{1*} \Omega_{+1,m} + \Omega_{-1,m}^{1*} \Omega_{+1,m-1} }{4\Delta_m}.
	\end{align}
\end{subequations}
Note that the coupling strengths are typically given as $g_{0,m}^l = g_{a_l} c_{m,m}$, $g_{\pm 1,m}^l = g_{b_l} c_{m,m\pm 1}$ with Clebsch-Gordan coefficients $c_{m,n}$ \cite{MetcalfBook} and thus all have a similar magnitude given by $g_{a_l}$ or $g_{b_l}$, respectively. Hence, if we also assume that $\Delta_m \approx \Delta_{m'}$ for all $m, m' = -j-1,\ldots,j+1$, the averages $|\delta^{+}_k|$ are typically larger than the deviations $|\zeta_{k,m}|/N$ from these averages.

To adiabatically eliminate the cavity modes in the regime, where the dynamics of the cavity alone is much faster than the coupled atom-cavity dynamics, we follow Ref.~\cite{Azouit2017}: For our model, the Liouville operator of the full system can be written as $\LL = \LL_\text{photons} + \eps \LL_\text{atoms} + \eps \LL_\text{int}$ with Liouville operators $\LL_\text{photons}$, $\LL_\text{atoms}$ and $\LL_\text{int}$ acting on just the photons, just the atoms and the joint system, respectively, and $\eps \gtrsim 0$ (here, e.g., $\eps = \left(\sum_k \delta_k^2 + \kappa_k^2\right)^{-1/2}$, after an appropriate rescaling of the time, $t\mapsto t/\eps$). Then the full density matrix $\rho_\text{full}$ and the time evolution of the reduced density matrix $\rho$ can be expanded as $\rho_\text{full} = \sum_{n=0}^{\infty}\eps^n\KK_n[\rho]$, $\dot{\rho} = \sum_{n=1}^{\infty} \eps^n \LL_n[\rho]$, where $\KK_n$, $\LL_n$ are linear time-independent superoperators. The zeroth order yields $\LL_\text{photons}[\KK_0[\rho]] = 0$, which is here solved by $\KK_0[\rho] = \rho \otimes \rho_\text{vac}$, where $\rho_\text{vac}$ is the vacuum state of the cavity. The first order gives
\begin{align}
	\nonumber
	\LL_1[\rho] &= \LL_\text{atoms}[\rho] + \Tr_\text{photons} \LL_\text{int}[\rho \otimes \rho_\text{vac}]\\
	&= \LL_\text{atoms}[\rho],\\
	\nonumber
	\LL_\text{photons}[\KK_1[\rho]] &= - \LL_\text{int}[\rho \otimes \rho_\text{vac}] \\
	&= - \i \! \sum_{k=a_1,a_2,b_1,b_2} \!\!\left(X_k^\dagger \rho \otimes k^\dagger \rho_\text{vac} - \rho X_k \otimes \rho_\text{vac} k \right).
	\label{eq:K1}
\end{align}
Equation~\eqref{eq:K1} can be solved towards $\KK_1$ using an ansatz $\KK_1[\rho] = \sum_{k=a_1,a_2,b_1,b_2} \left(\sigma_k X_k^\dagger \rho \otimes k^\dagger \rho_\text{vac} + \tau_k \rho X_k \otimes \rho_\text{vac} k \right)$, yielding coefficients $\sigma_k = \tau_k^* = (\i\kappa_k + \delta_k)/(\kappa_k^2 + \delta_k^2)$. Using that $\Tr_\text{photons} \KK_1[\LL_1[\rho]] = \Tr_\text{photons}\LL_\text{atoms}[\KK_1[\rho]] = 0$, the second order becomes
\begin{align}
	\nonumber
	\LL_2[\rho] ={}& \Tr_\text{photons} \LL_\text{int}[\KK_1[\rho]] \\
	\nonumber
	={}& \i \sum_{k=a_1,a_2,b_1,b_2} \frac{\delta_k}{\kappa_k^2 + \delta_k^2} \left[X_k X_k^\dagger, \rho\right] \\
	&+ \sum_{k=a_1,a_2,b_1,b_2} \frac{\kappa_k}{\kappa_k^2 + \delta_k^2} \left(2 X_k^\dagger \rho X_k - \left\{ X_k X_k^\dagger, \rho\right\}\right),
\end{align}
which together with the first order gives Eq.~\eqref{eq:finalResultExperiment} (after undoing the rescaling of time, $t\mapsto \eps t$).

\bibliographystyle{apsrev4-2}
\bibliography{library.bib}

\end{document}